\documentclass[aps,prl,reprint,twocolumn,showpacs,floatfix,superscriptaddress]{revtex4-1}

\usepackage{amssymb,amsmath,amstext}                
\usepackage{graphicx}                                               
\usepackage{epstopdf}                                               
\usepackage{color}                                                     
\usepackage{bm}                                                        
\usepackage{appendix}                                              
\usepackage[utf8]{inputenc}
\usepackage{bbold}
\usepackage{bbm}
\usepackage{latexsym}
\usepackage{xcolor}
\usepackage{braket}
\usepackage{ulem}
\definecolor{lblue} {RGB}{51,71,158}
\usepackage[colorlinks=true,citecolor=blue,linkcolor=blue,urlcolor=lblue]{hyperref}
\newcommand{\average}[1]{\overline{#1}}

\begin{document}

\title{Many-body localization in Bose-Hubbard model: evidence for the mobility edge}
\author{Ruixiao Yao}
\affiliation{School of Physics, Peking University, Beijing 100871, China}

\author{Jakub Zakrzewski}
\affiliation{Institute of Theoretical Physics, Jagiellonian University in Krakow,  \L{}ojasiewicza 11, 30-348 Krak\'ow, Poland }
\affiliation{Mark Kac Complex
Systems Research Center, Jagiellonian University in Krakow, Krak\'ow,
Poland. }
\email{jakub.zakrzewski@uj.edu.pl}

\date{\today}

                              
\begin{abstract}
Motivated by recent experiments on interacting bosons in quasi-one-dimensional optical lattice [Nature {\bf 573}, 385 (2019)] we analyse theoretically properties of the system in
the crossover between delocalized and localized regimes. Comparison of  time dynamics for  uniform and density wave like initial states enables demonstration of the existence of the mobility edge. To this end we define a new observable, the mean speed of transport at long times. It gives us an efficient estimate of the critical disorder for the crossover. We also
show that the mean velocity growth of occupation fluctuations close to the edges of the system carries the similar information. Using the quantum quench procedure we show that it
is possible to probe the mobility edge for different energies.
\end{abstract}

\maketitle
Many-body localization (MBL) despite numerous efforts of the last 15 years
(for some reviews see   \cite{Huse14, Nandkishore15, Alet18, Abanin19}) is still a phenomenon not fully understood. Recently even its very existence in the thermodynamic
limit has been questioned \cite{Suntais19} which created a vivid debate \cite{Abanin19z,Sierant19z,Panda19}. Simulations of large systems dynamics are also not fully conclusive \cite{Doggen18,Chanda19}. It seems that, as suggested by \cite{Panda19}, present day computer resources prevent us from drawing a definite conclusion on this point.  The issue may be addressed in experiments via quantum simulator approach \cite{Altman19sim} although the required precision may be also prohibitive. 

The experimental studies of MBL are much less numerous. Early work \cite{Kondov14} provided  indications of MBL  in a large fermionic system in an optical lattice. Subsequent studies 
showing the lack of thermalization and a memory of the initial configuration in time evolution were reported for interacting fermions in quasiperiodic disorder in optical quasi-one-dimensional (1D) lattice
\cite{Schreiber15}. Experiments also considered rather large systems with either fermions  \cite{Luschen17} or bosons \cite{Choi16} (the latter in two dimensions). Only recently quite small systems attracted experimental attention
first for interacting photons \cite{Roushan17}  where even the attempt at level spacing statistics measurement was made as well as for bosonic atoms where logarithmic spreading of entanglement was observed \cite{Lukin19} as well as long-range correlations analysed \cite{Rispoli19}. 

On the theory side bosonic system were not the prime choice for the analysis for a simple reason. While for spin-1/2 (or spinless fermion) the local Hilbert space dimension is two, twice that for spinful fermions, for bosons the strict limit is set by a total number of particles in the system. This severely limits the number of sites that can be included in any simulation
performed within exact diagonalization-type studies. In effect only few papers addressed MBL with bosons. By far the most notable is a courageous  attempt to treat bosonic system in two dimensions by approximate tensor network approach \cite{Wahl19} (still restricting the model to maximally double occupations of sites). Earlier treatment \cite{Sierant18} considered both small and large systems
in 1D revealing the existence of the so called reverse mobility edge in the energy spectra. With assumed 3/2 filling of the system it has been shown that higher lying in energy states are localized  for lower amplitude of the disorder for sufficiently strong interactions. {Many body localization with superconducting circuits were discussed in  \cite{Orel19} while the role of doublons for thermalization in \cite{Krause19}.} Very recently the same system was discussed also at half-filling \cite{Hopjan19} confirming the existence of the mobility edge. Let us mention for completeness also works on MBL of bosons with random interactions \cite{Sierant17,Sierant17b} instead of a random on-site potential. 

\begin{figure}
  \includegraphics[width=1\linewidth]{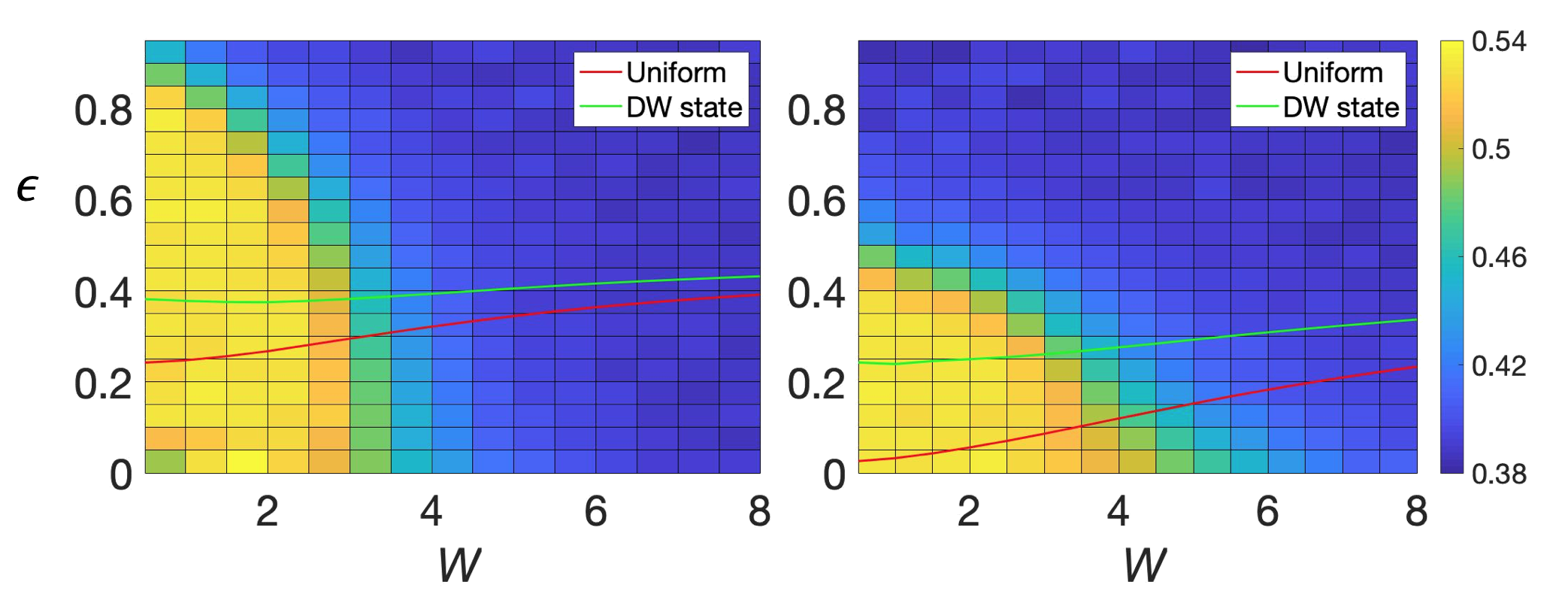}
  \vspace{-0.25cm}
  \caption{ {{Mean gap ratio $\average r$} as a function of the energy for $U=1$ (left) and $U=2.87$ right, for a system of 8 bosons on 8 lattice sites. Observe that for the higher $U$ case, high energy states are localized at lower disorder amplitudes revealing the inverted mobility edge.  Red and yellow lines represent the energy of the uniform initial state (with unit occupation of all sites) and the density wave with even states being doubly occupied while odd states being empty. Observe that for larger $U$ the transition to localized situation as revealed by $\bar r$ statistics depends strongly on the initial state energy, the density wave initial state should localize more easily.
  \label{fig:spacings} \vspace{-0.25cm}
 }}
\end{figure}
In these works the tight-binding model describing the physics is given by 1D Bose-Hubbard Hamiltonian:
\begin{equation}
H=-J\sum_i ^{L-1}\left ({\hat b^\dagger_i}{\hat b_{i+1}} +h.c.\right) +\frac{U}{2}\sum_i^L{\hat n_i}({\hat n_i}-1)+\sum_i^L\mu_i {\hat n_i},
\label{ham}
\end{equation}
where ${\hat b_i}$ (${\hat b^\dagger_i}$) are bosonic annihilation (creation) operators on site $i$ with $[{\hat b_i},{\hat b^\dagger_j}]=\delta_{ij}$ and ${\hat n_i}={\hat b^\dagger_i}{\hat b_i}$. We assume uniform interactions $U$ across the lattice while the on-site energies (chemical potential) $\mu_i$ is site dependent.   While \cite{Sierant18,Hopjan19} considered a random uniform on-site disorder, following the experiments \cite{Schreiber15,Lukin19,Rispoli19} from now on we assume the quasi-periodic disorder $\mu_i=W\cos(2\pi\beta i +\phi)$
where, for a given realization, $\phi$ is fixed and disorder averaging corresponds to an average over uniformly distributed $\phi\in[0,2\pi)$. It is known that the localization properties of the system strongly depend on the value of $\beta$ \cite{Guarrera07,Doggen19} , from now on  we take $\beta=1/1.618$ and a unit filling following \cite{Lukin19,Rispoli19} and work with open boundary conditions. {We consider sufficiently deep optical lattices, so \eqref{ham} may be used, for shallow quasiperiodic potential case see \cite{Hepeng19}.}

With the model defined we study first its spectral properties. The localization may be probed looking at the mean gap ratio, $\average r$ \cite{Oganesyan07} defined as an average of  $r_n$ - the minimum of the ratio of consequtive level spacings $s_n=E_{n+1}-E_n$:
\begin{equation}
r_n= \min \{ \frac{s_{n+1}}{s_n}, \frac{s_n}{s_{n+1}} \}.
\label{gap}
\end{equation}
$\average r$ is a simple, dimensionless probe of level statistics  \cite{Oganesyan07} with $\average r\approx 0.38$ for Poisson statistics (PS) (corresponding to localized, integrable case)
and $\average r\approx 0.53$ for 
 Gaussian orthogonal  ensemble (GOE) of random matrices well describing statistically the ergodic case
\cite{Mehtabook,Haake}. Diagonalizing the model for 8 bosons on 8 sites and calculating $\bar r$ as a function of the disorder amplitude $W$ and the relative energy $\epsilon_n=\frac{E_n-E_{min}}{E_{max}-E_{min}}$ we obtain the plot represented in Fig.~\ref{fig:spacings}. While for $U=1$ the energy dependence of $\bar r$ seems only weakly dependent on energy, for $U=2.87$ a clear inverted mobility edge emerges:  at the same disorder values while states of higher energies seem localized, the low lying states have $\bar r$ corresponding to extended states. This is in agreement with earlier studies \cite{Sierant18,Hopjan19} at  different densities.

\begin{figure}
\includegraphics[width=0.95\linewidth]{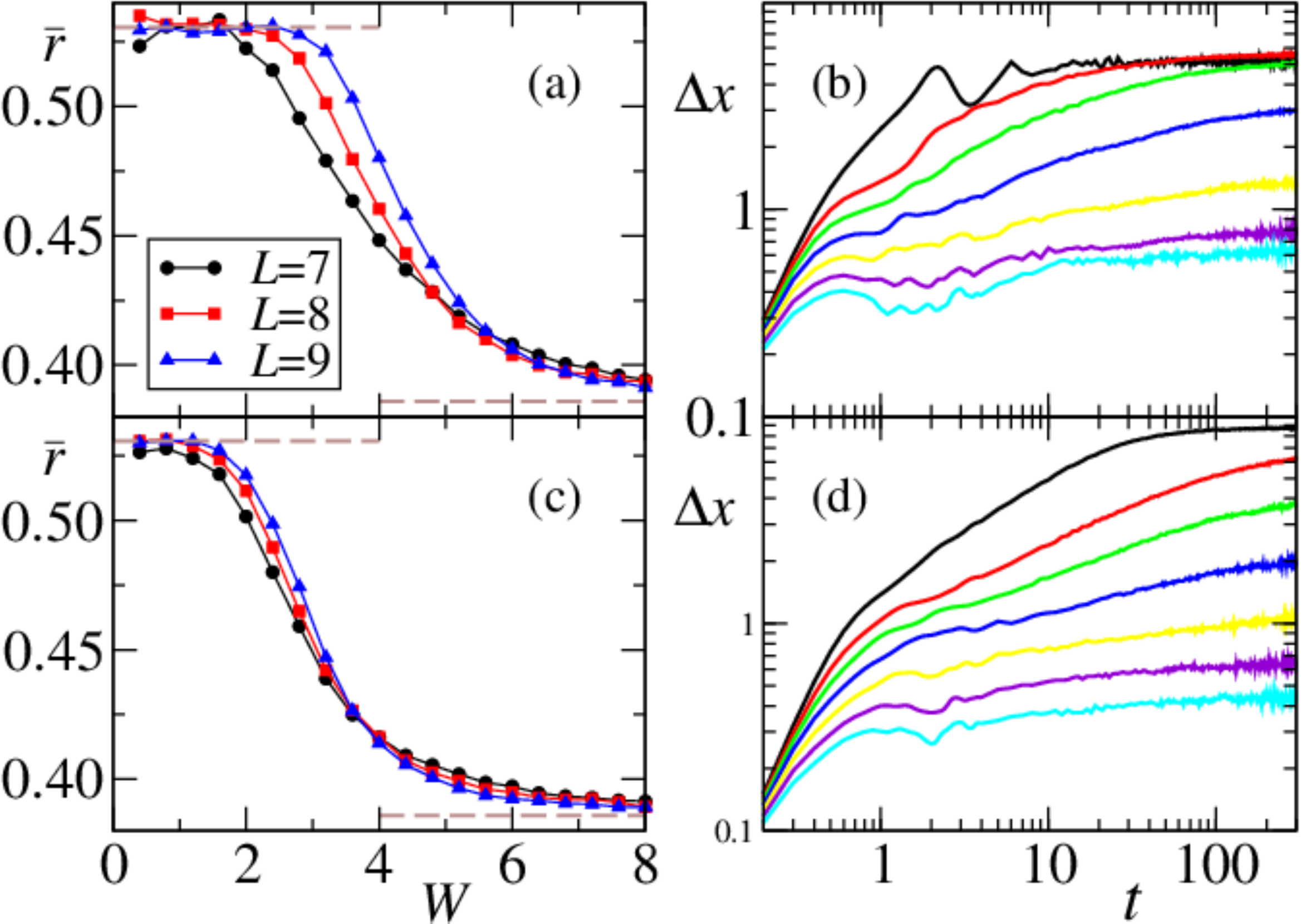}
  \caption{ Left: Mean gap ratio $\overline r$ reveals that transition to MBL, as indicated by crossing of curves for different system sizes (indicated in the figure) occurs for energies corresponding to a uniform state [(a) panel], $\epsilon\approx0.15$]  for much stronger disorder than for $\epsilon=0.3$ corresponding to the density wave initial state. This is manifested by the transport distance, $\Delta x$ (see text for the definition) as obtained from time dynamics starting from those two states  (right column) which reveals subdiffussive growth on the localized side (straight line behaviour in log-log plots with the slope well below 0.5). The curves correspond to $W=1,3,4,5,6,7,8$ from black (top) to cyan (bottom). The rapid ballistic-like growth at $W=1$ in both cases saturates at intermediate times due to a finite system size - the time dependence is shown for $L=8$.   
  \label{newfig2}
 }
\end{figure}
A  more qualitative analysis might be carried out with finite size scaling (FSS) \cite{Luitz15}. Recent works have indicated \cite{Mace19,Laflorencie20,Suntajs20} , however, that MBL transition may be of Kosterlitz-Thouless type rendering symmetric FSS questionable. In view of that we show  $\overline r$ as a function of disorder for different system sizes with crossing of the curves being an indicator of the crossover disorder value $W_c$ - compare Fig.~\ref{newfig2}. We may observe that the transition for $U=2.87$ clearly depends on the initial state energy suggesting the existence of the mobility edge. The crossing between $L=8$ and $L=9$ data occurs for $W_c=6.5$ ($W_c=4.0$) for $\epsilon$ corresponding to uniform - UN (density wave - DW) state. On the other hand for $U=1$ $W_c\approx4.6$ for both UN and DW  energies (see \cite{suppl} for the plot, there also FSS is discussed further).

We now pose a question - can the mobility edge manifest itself in experiments? Since individual levels are hardly accessible, we shall seek manifestations of the mobility edge in observables accessible to measurements in time dynamics. Instead of the imbalance, possible to be used for different DW type states \cite{Sierant18}   we shall consider  the transport distance (for other studies of the related quantity for spin systems see \cite{Bera17,Weiner19}) defined as \cite{Rispoli19}: $\Delta x=2|\sum_d d\times\overline{\langle G_c^{(2)}(i,i+d)\rangle_i}|$ - i.e. the modulus of  the disorder averaged (as denoted by the overbar) and site-averaged (as denoted by a subscript $i$) second order correlation function of density $G_c^{(2)}(i,i+d)= \langle \hat n_i\hat n_{i+d}\rangle - \langle \hat n_i\rangle\langle \hat n_{i+d}\rangle$. We concentrate mostly on $U=2.87$ case which reveals mobility edge in above spectral analysis and confront them with data obtained for $U=1$ when no mobility edge evidence is expected for UN or DW  states. 
 
The time dependence of the transport distance obtained for $L=8$ system is shown in Fig.~\ref{newfig2} in right panels for different values of the disorder. Data are obtained using exact diagonalization approach. On the delocalized side one observes a fast almost ballistic growth (for $W=1$) followed by a saturation when the finite size of the system dominates the dynamics. Upon approaching the transition the power-law growth becomes apparent for intermediate times, the corresponding power $\beta=d \ln\Delta x/d\ln t$ decreasing smoothly within the interval of [0,0.35] indicating a subdiffusive dynamics for  both initial states considered (similar behavior is observed for $U=1$ \cite{suppl}). For sufficiently large $W$ the motion freezes with very small $\beta$. The data are averaged over 200 disorder realizations, the residual fluctuations apparent in the figures form an estimate of the statistical error.
\begin{figure}
\includegraphics[width=0.9\linewidth]{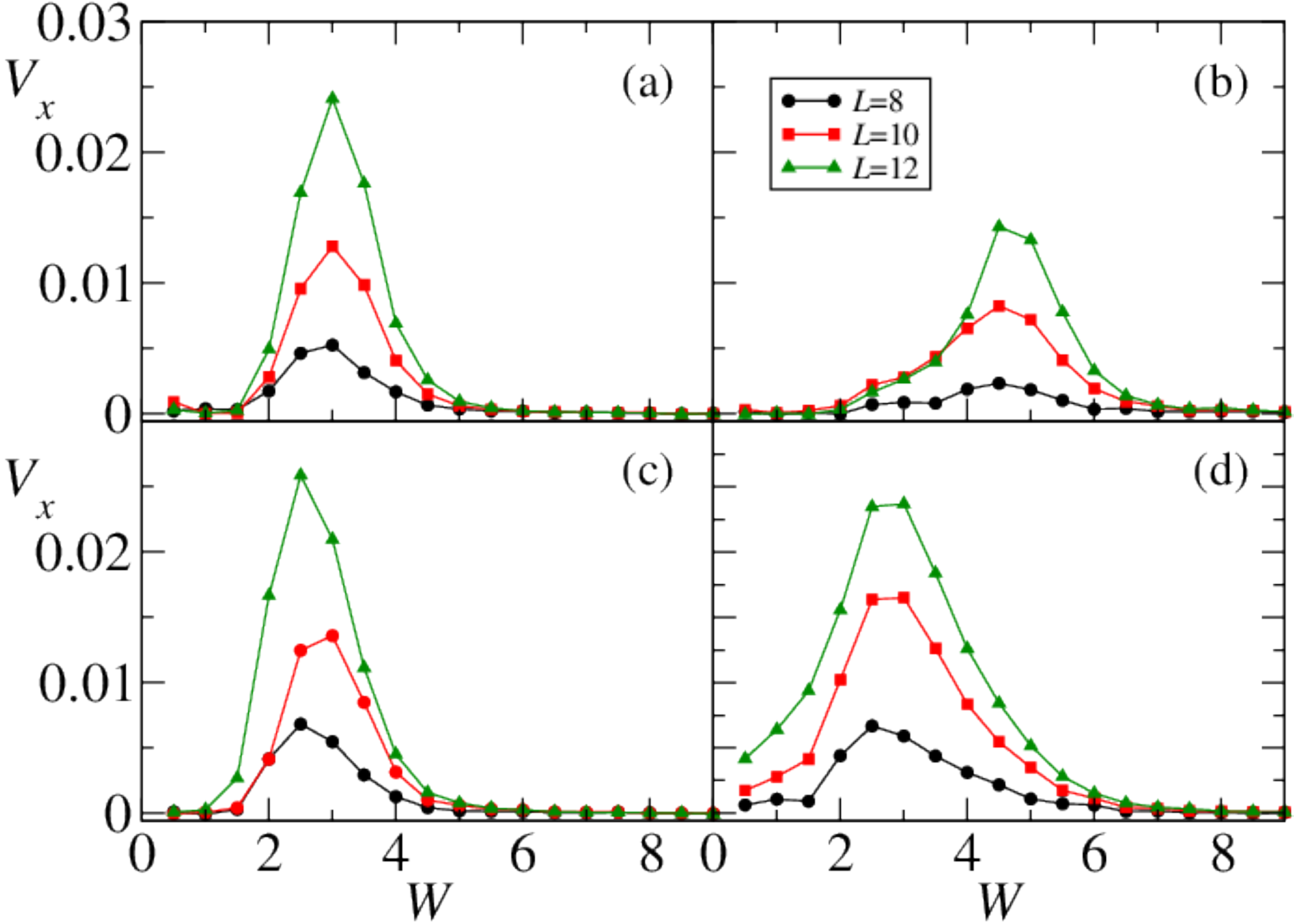}
  \vspace{-0.25cm}
  \caption{ {The mean transport velocity $V_x$ (averaged in $t\in[100,250]$ interval) as a function of the disorder for $U=1$ (left) and $U=2.87$ (right) for UN state (top) and DW (bottom). The derivative seems to indicate the transition to MBL quite clearly, again manifesting different critical disorder value (when it vanishes on the localized side)  for UN and for  DW state at $U=2.87$. Statistical errors of $V_x$ are smaller than the symbols. 
    \label{fig:derfin} \vspace{-0.25cm}
 }}
\end{figure}

Let us inspect in more detail the system properties in subdiffusive region.  Apart from $L=8$ case, we consider also $L=10,12$ to observe the effects due to the system size. For $L=10,12$ time evolution is carried our using the standard Chebyshev technique \cite{Kosloff84,Fehske08}. Due to a moderate number of 200 disorder realizations 
to minimize local fluctuations one may  average $\Delta x(t)$ obtained over some time  interval, for sufficiently large times. On the localized side one expects such an average, $\Delta x_L$
to be small, signalling the transition. While such an approach signals  the mobility edge for $U=2.87$ (see\cite{suppl}), a more dramatic effect is observed taking 
the mean time derivative of $\Delta x(t)$ over some large interval.  Such velocity of transport $V_x$ can be readily obtained as  $V_x=[\Delta x(t_2)-\Delta x((t_1)]/(t_2-t_1)$. It is shown for $t_1=100$ and $t_2=250$  in Fig.~\ref{fig:derfin}. The larger the system size the more pronounced are the maxima of the derivative.  On the delocalized side $V_x$ is small as $\Delta x$ saturates due to  finite sizes considered. On the localized side $V_x$ practically vanishes as transport for long times is prohibited. Thus the pronounced maximum of the derivative in the critical transition regime is a robust and {\it aposteriori} expected phenomenon. The critical disorder is read out as the point where $V_x$ practically vanishes (being say 1\% of the maximal value). While it is important to analyse $V_x$ at large times, after the initial growth, the results are robust against varying the choice of the time interval as well as the method of estimating the mean velocity \cite{suppl}. 
For $U=1$ (left column) the uniform (a) and the DW (c) initial state lead to similar $V_x$ dependence as a function of $W$.
$V_x$ approaches 0  and becomes almost independent of the system size for
$W\approx 5\pm 0.5$ which nicely matches the critical disorder values obtained from gap ratio analysis. 

For $U=2.87$ case, the position of the peaks of $V_x$ for UN (b) and DW (d)  states in Fig.~\ref{fig:derfin} strongly differ.  Using the same criterion of vanishing $V_x$ in the localized regime, we get the critical disorder  $W_c\approx7$ for UN state and $W_c\approx 5.5$ for the DW (the latter value being a bit too large with respect to gap ratio data).
\begin{figure}
\includegraphics[width=0.9\linewidth]{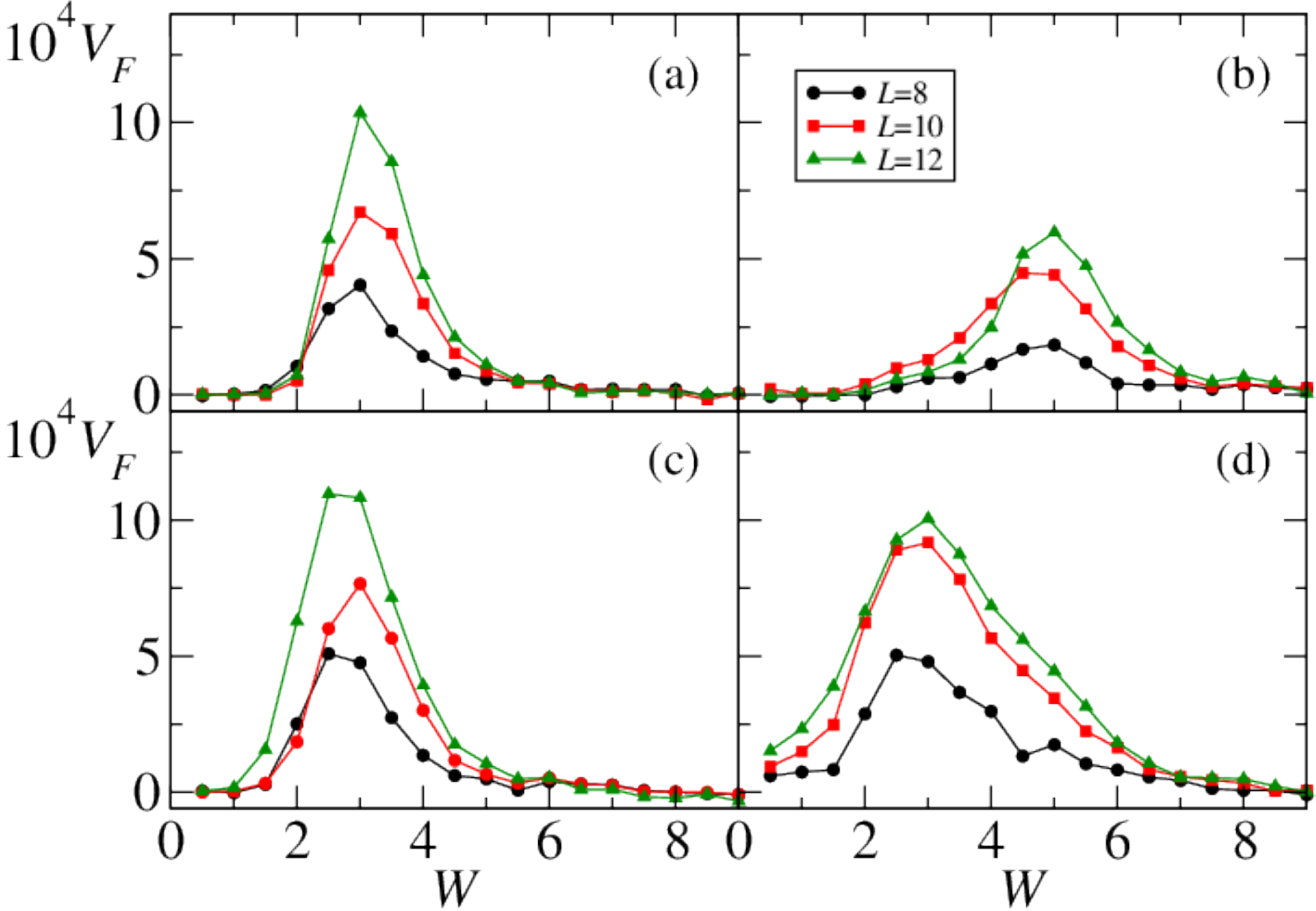}
  \vspace{-0.25cm}
  \caption{ {Same as Fig.~\ref{fig:derfin} but for the edge fluctuation velocity $V_F$. For $U=2.87$ (right) a clear difference between UN  and DW initial states also appears, the transition to MBL occurring for larger $W$ for UN state, in agreement with the gap ratio analysis. Statistical errors in $V_F$ determination are comparable to symbols' sizes.
    \label{fig:derfluc} \vspace{-0.25cm}
 }}
\end{figure}

While the evidence for the mobility edge existence apparent in Fig.~\ref{fig:derfin}  is quite strong, determination of $\Delta x$ requires measurements of all second order correlations. As it turns out this may not be necessary. Consider on-site fluctuations  $F_i = \langle n_i^2\rangle - \langle n_i\rangle^2$. While \cite{Naldesi16} considered fluctuations in the middle of the chain, it is advantageous to concentrate on $F=(F_1+F_L)/2$ i.e. averaged fluctuation on the edges. This follows the argument that edges are least sensitive to the system size \cite{Khemani17}. While already the fluctuation, $F$, at long times shows indications of the crossover to localized phase \cite{suppl}, we present in   Fig.~\ref{fig:derfluc} its derivative $V_F=dF/dx$ averaged over ``long times'' in a manner completely analogous to $V_x$. Remarkably, it also reveals
the mobility edge when varying $W$ in a similar manner to $V_x$ plotted just above.

\begin{figure}
\includegraphics[width=1\linewidth]{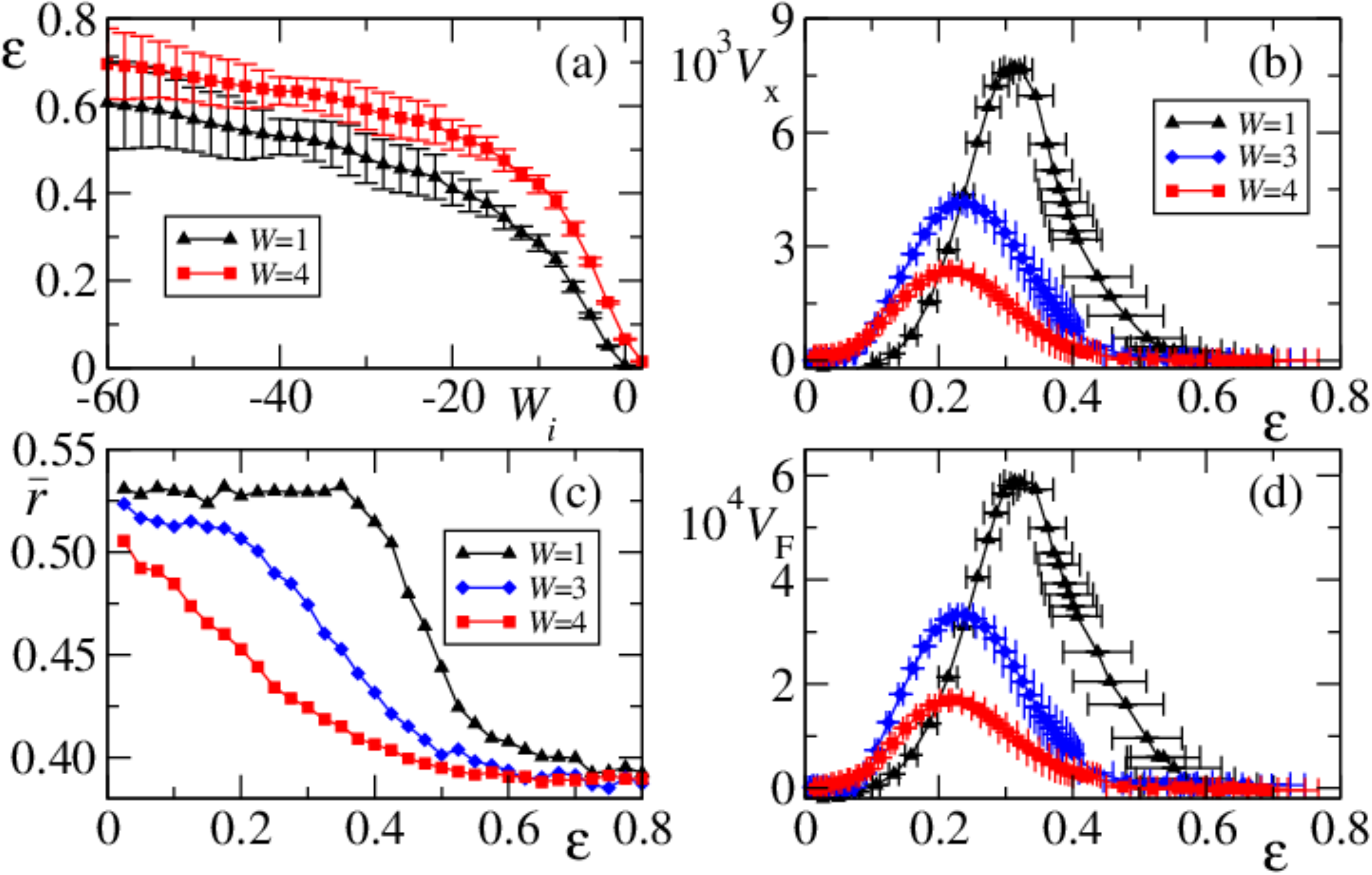}
  \caption{ {(a): Quantum quench from initial disorder amplitude $W_i$ to final values $W$ (as indicated in the figures) allows to prepare initial states of different energy $\epsilon$.
  (b): The mean transport velocity at long times $V_s$ obtained for such prepared states reveals a  $W$-dependent crossover to MBL  as a function of $\epsilon$ visualizing the mobility edge seen in Fig.~\ref{fig:spacings}(b). The crossover $epsilon$ values agree with gap ratio statistics for corresponding $W$ values shown in panel (c). Edge-fluctuation velocity $V_F$ [panel (d)] is also sensitive to the crossover in a manner similar to $V_s$.   
    \label{fig:quench}
 }}
\end{figure}
Our analysis up till now have been restricted to initial Fock-like states and the corresponding energies. It was suggested, however, that different energy regimes may be addressed by a ``quantum quench spectroscopy'' \cite{Naldesi16} to reveal the mobility edge energy dependence. The proposed scheme assumes a preparation of the ground state of the system at some value of a parameter, then  the fast quench of that parameter to the final investigated value transfers the system into an excited wavepacket with excess energy dependent on the change of this parameter. By changing the initial parameter value one may, hopefully, scan the final energy. This method was tested
in \cite{Naldesi16} on spinless fermions case with quenching the disorder value and observing the time dependence of the entanglement entropy as well as on site density fluctuations.

 Let us apply the same method to the current problem. We assume that the ground state is prepared at different disorder amplitudes $W_i$  and  the rapid quench brings the disorder amplitude to a desired final value $W$. Changing $W_i$ we may  change the energy ($\epsilon$) of the prepared wavepacket as shown in  Fig.~\ref{fig:quench}(a). Note that for reaching significant excitation $\epsilon$ we start with ``negative'' amplitude $W_i$. In this way we can reach different final $W$ values. The obtained energies are characterized by, unfortunately, quite significant error. For a given prepared initial $W_i$ different realizations of the disorder (different phase $\phi$) lead to different excitation energies. 
 As noted in \cite{Naldesi16} the resulting error diminishes with the increasing system size, here we consider a small system $L=8$ mostly (see \cite{suppl} for larger systems) as 
 then gap ratio statistics allows us to verify the results obtained. The error bars shown in Fig.~\ref{fig:quench} are due to this effect and limit the accuracy of the final energy obtained by quenching. 
 
 The initial state is then evolved in time determining  $\Delta x(t)$ and its mean derivative $V_x=d\Delta x/dt$ exactly as before. At different final disorder values a transition from delocalized to
localized regimes is observed when changing ``smoothly'' $\epsilon$ -- compare Fig.~\ref{fig:quench}(b). Again the tail of $V_x$, when its value becomes close to zero is an indicator of MBL
regime. For comparison,   Fig.~\ref{fig:quench}(c) shows the corresponding  gap ratio statistics at these disorder values. The agreement is spectacular showing that the quantum quench spectroscopy combined with the transport distance measurements allows to continuously monitor the mobility edge in our system. The drawback of the method is the inherent limitation of the energy $epsilon$ resolution discussed above.

To complete the picture, Fig.~\ref{fig:quench}(d) presents the edge-fluctuation derivative $V_F=dF/dx$ averaged over ``long times''. As for UN and DW states in Fig.~\ref{fig:derfluc} it also reveals the mobility edge when varying final disorder strength $W$. This is quite promising for possible applications of quench spectroscopy to systems of larger size where measurements leading  to transport distance (and $V_x$) may become costly while measurements of edge fluctuations require site resolution at the edges only.  

To conclude, by considering transport properties in the transition between extended and localized states in Bose-Hubbard Hamiltonian describing bosons in optical lattice with diagonal quasiperiodic disorder we have shown that observables directly accessible to the experiment \cite{Rispoli19} reveal the existence of the mobility edge in the system. The mobility edge has been convincingly shown to exits  for the finite size Heisenberg chain \cite{Luitz15} studying different spectra measures  (most notably the gap ratio statistics) as well as using quantum quench spectroscopy and time dependence of the entanglement and number fluctuations \cite{Naldesi16}. It has also been 
postulated to exist for Bose-Hubbard system with a larger density by considering different density wave states \cite{Sierant18}. Here we show that within the realm of state of the art, current experiment \cite{Rispoli19} the mobility edge may be verified, for the first time, experimentally, via readily accessible observables. We considered different initial states, either prepared as Fock states with different density patterns or states obtained via quantum quench of the disorder amplitude. The latter approach allows us (modulo inherent uncertainties) to follow the mobility edge in energy. Both the global trends and values of the critical disorder obtained by us quantitatively agree across different observables taken for analysis and agree with the spectral  gap ratio analysis.

{When this work was completed we have learnt that energy resolved MBL was also considered in spin quantum simulator \cite{Guo19}.}

 \begin{acknowledgments} 
 We are grateful to Sooshin Kim and Markus Greiner for discussing the details of simulations accompanying the experiment \cite{Rispoli19} {and to Piotr Sierant for remarks on the manuscript}.
JZ thanks Dominique Delande and Titas Chanda for
discussions. 
The numerical computations have been possible thanks to  High-Performance Computing Platform of Peking University. Support of PL-Grid infrastructure is also acknowledged.
This research has been supported by 
 National Science Centre (Poland) under project  2016/21/B/ST2/01086 (J.Z.). 
 \end{acknowledgments}

%


\section{Supplementary material}

\newcommand{\snum}{S}

\renewcommand{\theequation}{\snum.\arabic{equation}}
\renewcommand\thefigure{\snum.\arabic{figure}} 
\setcounter{equation}{0}
\setcounter{figure}{0}


\subsection{Comments on gap ratio analysis and finite size scaling}

\begin{figure}[h!]
\includegraphics[width=0.9\linewidth]{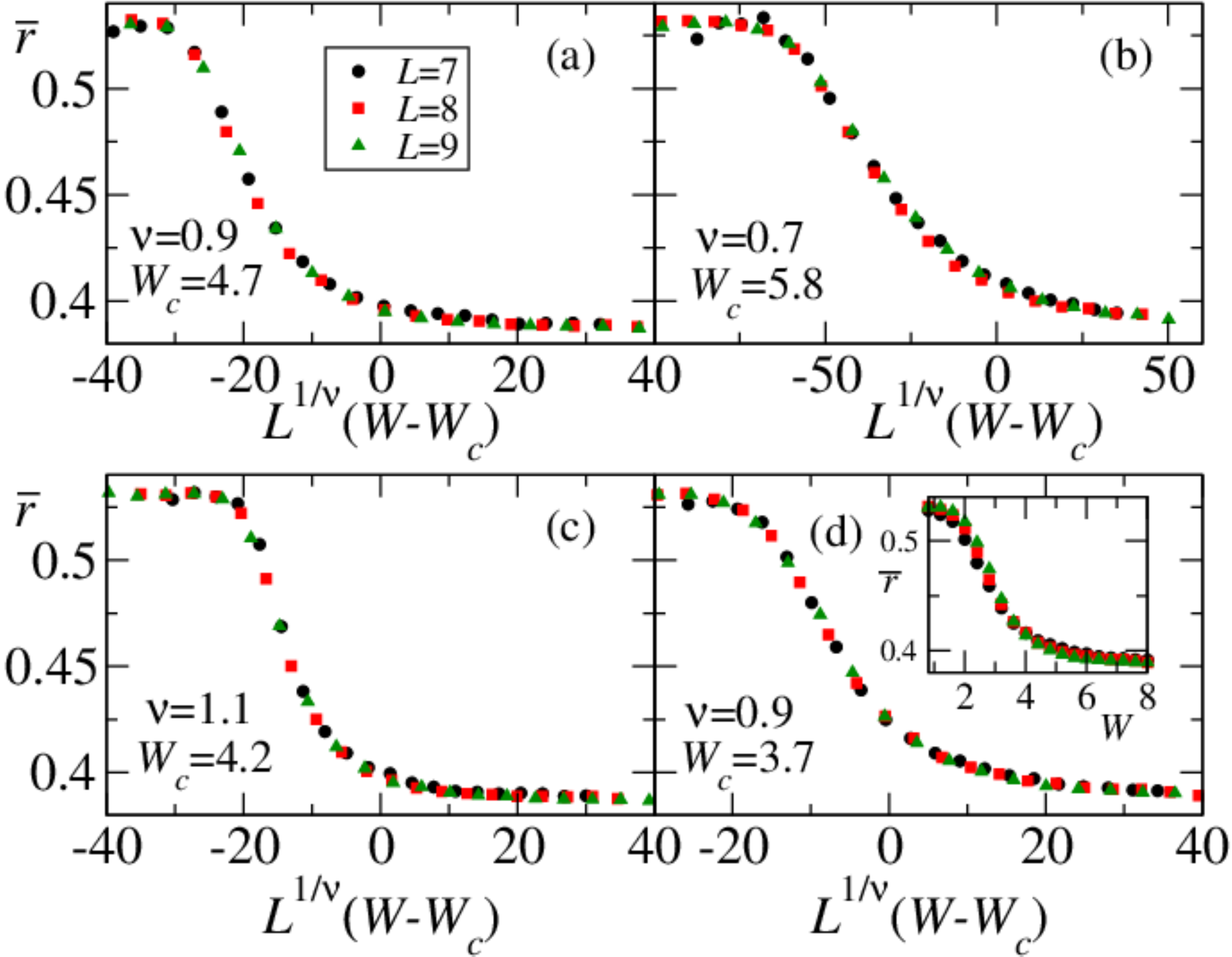}
  \caption{{1 (left) and U = 2.87 (right) in the scaled energy regions corresponding to two interesting states. Top row corresponds to$ \epsilon = 0.30/0.15$ (left/right) energy region appropriate for uniform initial state for L = 8 (compare to Fig. 1 in main text). Bottom row shows scaling for $\epsilon = 0.40/0.25$ (left/right), respectively values, appropriate for density wave initial state (for L = 8). The obtained values of critical disorder amplitude and critical exponent are indicated in the figure. The inset shows unscaled data for comparison. Data are obtained for 2000 disorder realizations.}}
  \label{FSS}
\end{figure}

In the letter we did not show the results of finite size scaling analysis. While several authors used a simple, single parameter scaling of the form $W \rightarrow (W - W_c)L^{1/\nu}$ 
to extract the critical disorder value $W_c$, this procedure has been put recently under critique. Already \cite{Luitz15} noticed that the obtained values of $W_c$ and $\nu$
dangerously depend on system sizes taken for the analysis. Moreover, the exponent $\nu$ appears to be close to unity which violates the so called Harris bound (for details see 
\cite{Luitz15} and references therein). Having at our disposal systems 
 with $L = 7,8,9$ we may perform such an FSS at energies corresponding to uniform (UN) and density wave (DW) states considered - the results are 
  shown in Fig.~\ref{FSS}. The difference in $W_c$ for U = 2.87 for $\epsilon$ values corresponding to UN and DW states is significant - giving an argument towards the existence of the mobility edge. On the other hand variations of $\nu$ are large (20\%) while there is no reason to believe that the character of the transition changes. Moreover, $\nu$ values, as for spin systems, violate the Harris bound. 
  
 Recent works on spin systems tend to describe MBL transition as being of Kosterlitz-Thouless (KT) type \cite{Laflorencie20,Suntajs20} and perform FSS with appropriate KT correlation function. We refrain from doing so as for bosons much smaller system sizes are available and such a procedure would also be doubtful. For that reason we use a simple crossing of $\overline r$ curves for different system sizes as a reasonable estimate of the characteristic disorder when MBL sets in. One should also keep in mind that for such small sizes (as also used in the experiment \cite{Rispoli19}) the crossover region must take a finite range of disorder amplitudes, $W$, so determining a precise number for $W_c$ in the thermodynamic limit is not needed really.

 \subsection{ Details on accuracy issues and numerical simulations}

Both spectral data (gap ratio) and time dynamics are obtained with ``exact'' methods such as an exact diagonalization and/or Chebyshev propagation scheme
(for sizes $L=10,12$). Thus the errors for the quantities used in the text come from two factors. Firstly, the results are averaged over the different realization of the 
quasiperiodic potential via drawing the phase $\phi$ from a random uniform distribution from $[0,2\pi]$ interval. Those errors may be minimized by simply increasing the number of
disorder realizations. Secondly, we define ``average'' or ``mean'' quantities at long times such as the mean transport distance, the mean derivative of the transport distance etc.
which are the quantities averaged over some time interval. These definitions inherently induce additional errors that we describe below.

\begin{figure}[h!]
\includegraphics[width=0.9\linewidth]{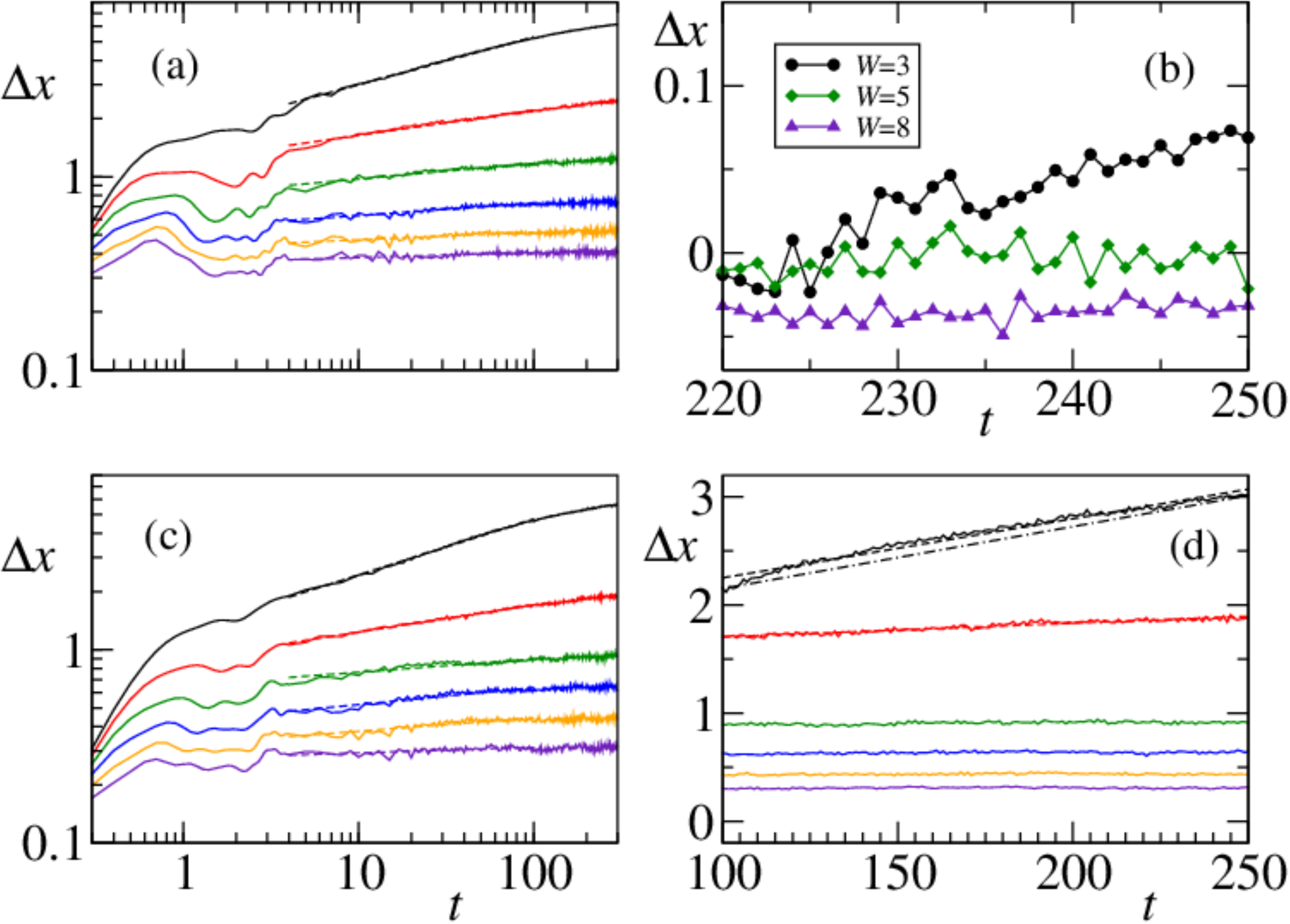}
  \caption{ {Left: The transport distance $\Delta x$ for $U=1$ case for a uniform density (a) and the density wave (c) states for $W=3,4,5,6,7,8$ values from the top to the bottom.
  Notice almost identical behaviour of the corresponding curves indicating that the crossover to MBL for both states occurs at approximately same values of $W$. The dashed lines
  show a linear fit (in the log-log plot) indicating a power law time-dependence for intermediate times. Panel (c) enlarges (in linear scale) the region of panel (a) used to determine ``the mean transport distance $\Delta x_L$ (see text). Different curves are shifted vertically for a better comparison. The fluctuations observed are due to a finite 200 disorder realizations (discrete points correspond to the finite step of the output from the numerical code. Panel (d) visualizes errors in the mean velocity of the transport $V_x$ - obtained as a mean slope in [100,250]
  interval. Dashed line corresponds to a linear fit, dashed-dotted line connects simply the $t=100$ and $t=250$ points for $W=3$. As discussed further in the text both can be used to estimate $V_x$.
    \label{deltasup} 
     }}
\end{figure}
Let us recall from the main text the definition of the transport distance    $\Delta x=2|\sum_d d\times\overline{\langle G_c^{(2)}(i,i+d)\rangle_i}|$ \cite{Rispoli19}. Here the second order correlation function of density $G_c^{(2)}(i,i+d)= \langle \hat n_i\hat n_{i+d}\rangle - \langle \hat n_i\rangle\langle \hat n_{i+d}\rangle$ is averaged over disorder realizations and different sites $i$. As shown in Fig.~2 of the letter, compare also Fig.~\ref{deltasup} the transport distance first rapidly grows (on the time scale of few tunnelings) redistributing particles. On a longer time scale two main features are observed. For small disorder (delocalized regime) $\Delta x$ saturates due to a finite small size of the system. With increasing disorder this saturation shifts to longer times, see $W=3$ curves which bends indicating a saturation at times $t>300$. Before reaching this stage the growth of $\Delta x$ with time follows to a good accuracy a power law behaviour (as shown by dashed lines in left panels of Fig.~\ref{deltasup}) with the power $\beta \in (0,0.35)$.  Both uniform and density wave initial states
in Fig.~\ref{deltasup}, panel (a) and (c), correspondingly, behave almost identically for the same $W$ values - showing the same effect of disorder. On a localized side both the growth and the values reached are quite small allowing for identification of the localized regime.

\begin{figure}
\includegraphics[width=0.9\linewidth]{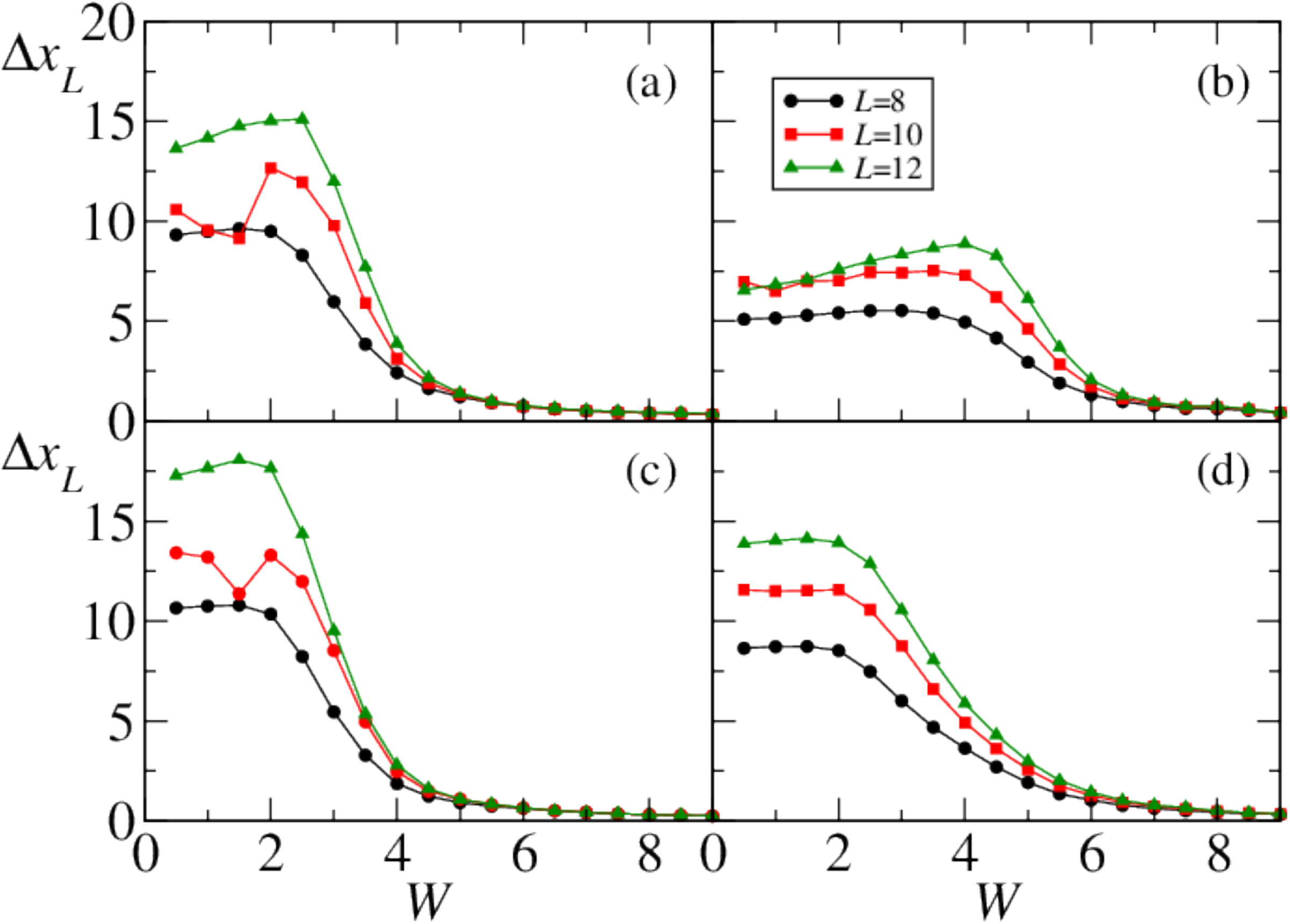}
  \vspace{-0.25cm}
  \caption{ {The transport distance $\Delta x_L$ obtained around $t=250$ as a function of the disorder for all cases studied (the order of panels is as in the previous figure). Data are for different system sizes. Observe that for $U=1$ (left column) the data for uniform and DW initial states are similar. For $U=2.87$ (right panel) a clear difference between uniform (top, (b)) and DW (bottom (d)) states appear, the transition to MBL occuring for large $W$ for the uniform state in agreement with gap ratio analysis.
    \label{fig:dxfin} \vspace{-0.25cm}
 }}
\end{figure}
One could attempt to use these log-log fits to estimate the transition to MBL (e.g. when $\beta$ decays to zero). Such a procedure has many drawbacks, as regions of approximate power law growth change with $W$, the smaller $\beta$ the errors become more important. Instead, also in view of short time-scale fluctuations due to a finite number of disorder realizations, it may be useful to consider   
a mean transport distance, $\Delta x_L$ at some experimentally reachable time of few hundreds of tunneling times. We find the mean value of the transport distance in the interval $t\in[220,250]$. Those data are shown in Fig.~\ref{fig:dxfin}. The numerical values depend of course on the chosen initial and final times (here 220 and 250) since apart from strongly localized regime the mean distance still grows (compare Fig.~\ref{deltasup}(c)). This dos not affect the determination of the critical disorder for the transition as we
define it as the place where  $\Delta x$ practically vanishes (reaches, say, 1/1000 of the maximal value) - then the error of  $\Delta x_L$ is very small. For small disorder values,
e.g. $W=3$ the error comes not from the disorder induced fluctuations (which are the main reason for averaging over a finite time interval) but from the remaining growth. Still this
error for $W=3$ which may be estimated from  Fig.~\ref{deltasup}(b) to be about 0.05 is less than one percent of the mean value in this interval $\Delta x_L\approx 6.1$ [compare
Fig.~\ref{fig:dxfin}(a)]. Note that  in Fig.~\ref{deltasup}(c) data are shifted vertically so behavior at different $W$ may be compared and fluctuations visualized. 

The main observable we use to identify the critical disorder strength for a given energy is, however, not the mean distance but rather the mean velocity of transport $V_x$ for intermediate and large times. For two times $t_1$ and $t_2>t_1$ it can be defined as $V_x=[\Delta x(t_2)-\Delta x((t_1)]/(t_2-t_1)$ and it is nothing else as ${\mathrm d} \Delta x(t)/ {\mathrm d} t$ averaged over $[t_1,t_2]$ interval. Alternatively one may fit a linear slope to $\Delta x(t)$ in the same interval. Both procedures are illustrated in Fig.~\ref{deltasup}(d)
for $W=3$ curve and lead, practically, for a sufficiently large interval, to almost identical slopes. The error of of the slope fitting procedure in $[100,250]$ interval is of the order of 1\% at most with errors being of the size of the symbols.   

\begin{figure}[h!]
\includegraphics[width=0.9\linewidth]{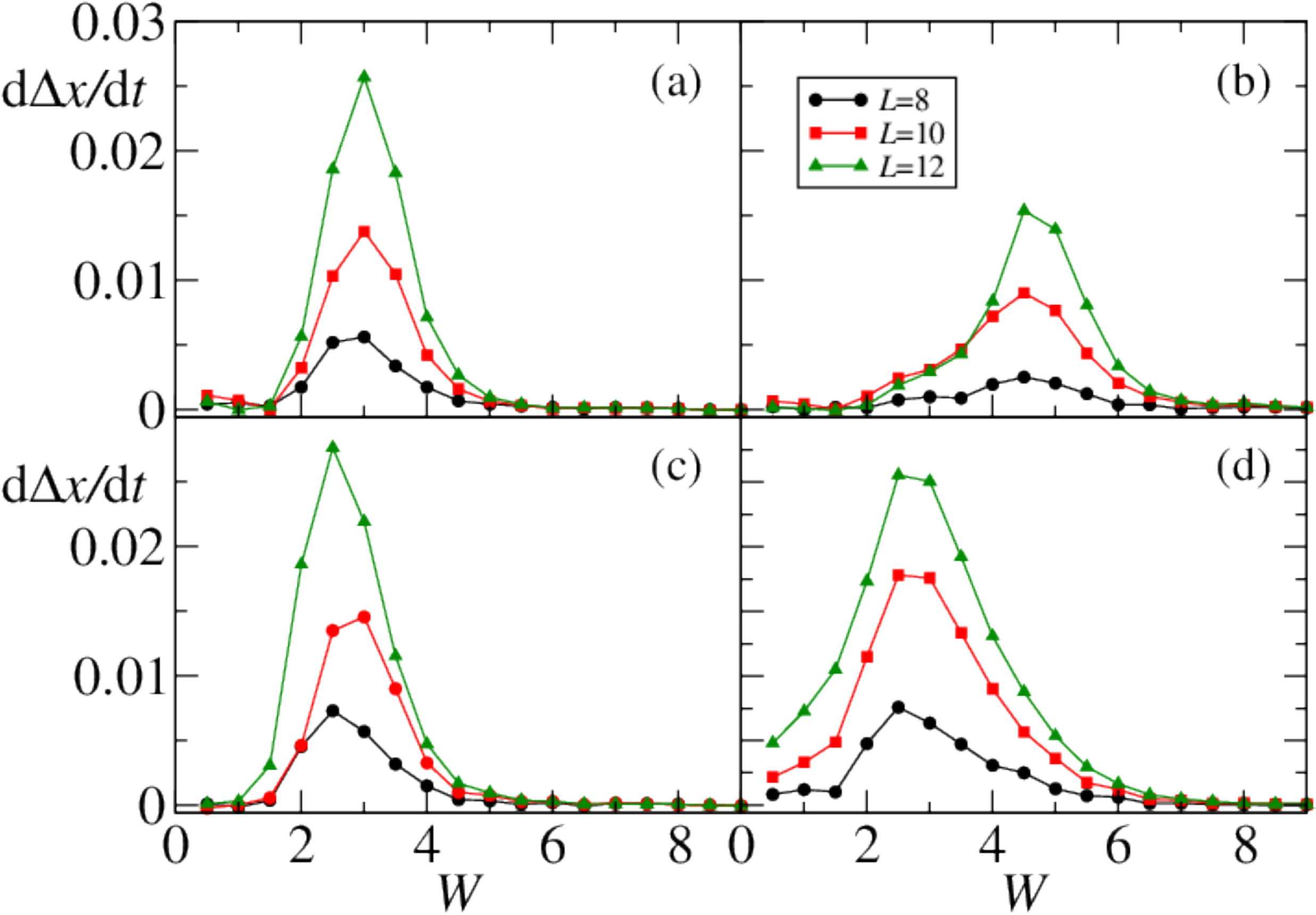}

  \caption{ {The mean derivative of the transport distance $V_x$ (averaged in $t\in[120,200]$ interval) as a function of the disorder for all four cases studied. 
  The data agree well with those of Fig.~4 in the main text despite a different time interval being taken for analysis. The almost identical points are obtained using the wieghted velocity average procedure - see text.
    \label{derwei} 
 }}
\end{figure}
The choice of the interval $[t_1,t_2]$ is not important, as long as we consider the interval for $t_1$ sufficiently large, say $t_1>80$. The mean velocity obtained 
for $[120,200]$ time interval are shown in Fig.~\ref{derwei}. Also,
instead of a linear fit of $\Delta x(t)$ with equal weights we may calculate the weighted linear fit in which squared deviations are weighted
 by a gaussian  centered at the center of the interval considered with a standard deviation $\sigma=25$. Transport velocities obtained  by such a procedure for 
the center at $t=175$ in the same interval yield essentially the same data as in Fig.~\ref{derwei} i.e. within the size of the symbols. This indicates that long time mean transport velocity is a robust measure.

\subsection{Quantum quench --additional results}

Here we supplement the results presented in Fig.~5 of the main text for the quantum quench scenario.
  
\begin{figure}[h!]
\includegraphics[width=0.9\linewidth]{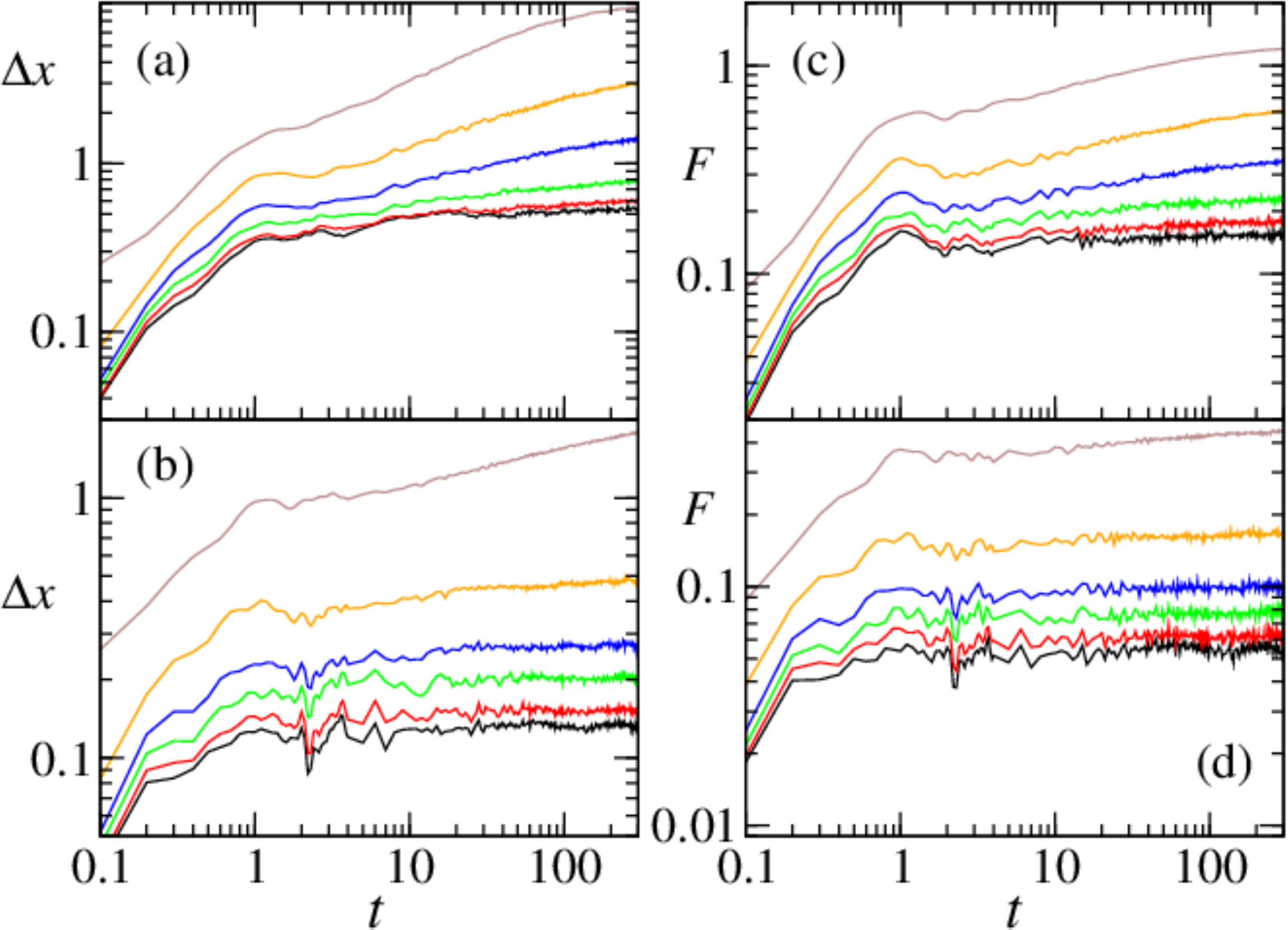}
 \caption{ {Transport distance $\Delta x$ (left: panels (a-b)) and the edge-fluctuation $F$ (right (panels (c-d)) as a function of time for   $W=1$ (top) and $W=3$ bottom.
  Different curves correspond to different initial starting disorder amplitude $W_i$. The curves are naturally ordered by $W_i=-60,-50,-40,-30,-20,-10$ from the bottom as the biggest $|W_i$ corresponds to highest excitation energies $\epsilon$ that are in the localized regime. Observe clear differences between upper and lower curves indicating the appearance of the mobility edge as the transition for $W=1$ occurs for different $W_i$ (and thus different $\epsilon$) than for $W=3$. 
    \label{timexf}
 }}
\end{figure}
In the quantum quench scenario \cite{Naldesi16} applied in the letter the system is prepared in the ground state at some disorder amplitude $W_i$ which is at t=0 switched abruptly to the desired $W$ value. Thus at $t=0$ the system is prepared in some wavepacket of energy $\epsilon$ (using scaled units as defined in the main text), the value of $\epsilon$ being dependent on the initial $W_i$ and the particular disorder realization. Scanning $W_i$ allows to scan $\epsilon$ as shown in Fig.~5(a) for L=8 and in this way realize a ``vertical cut'' of Fig.~1 at some $W$ value. Time evolving this state we may measure, like for initial Fock-like states the transport distance $\Delta x(t)$ as well as fluctuations on edges $F(t)$. Those are presented in. Fig~\ref{timexf} for two exemplary  disorder values (and several initial $W_i$). Since the initial state is not strictly separable, $\Delta x(0)$ does not strictly vanish, however, it is very small as amplitudes $W_i$ needed
to excite the system must be quite large (in absolute values).As seen in left panels of Fig.~\ref{timexf} the evolution resembles that for Fock states with the initial rapid growth on the scale of the tunnelling time and the power law sub-diffusive growth for larger times. To probe a broad range of excitation energies it turns out that $W_i$ should have an opposite sign to $W$ assumed for time evolution, the smaller $W-W_i$ the smaller $\epsilon$. Looking back at Fig.~1 localized states correspond to large $\epsilon$ - thus large $|W_i|$ and for those cases the growth of $\Delta x(t)$ practically stops. The significant difference between $W=1$ and $W=3$ curves is an another indicator of the mobility edge. Note that edge fluctuations (right panels) behave in quite a similar manner, thus site-fluctuations bring similar information to the transport distance (but does not require two-point correlations). 

\begin{figure}[h!]
\includegraphics[width=0.9\linewidth]{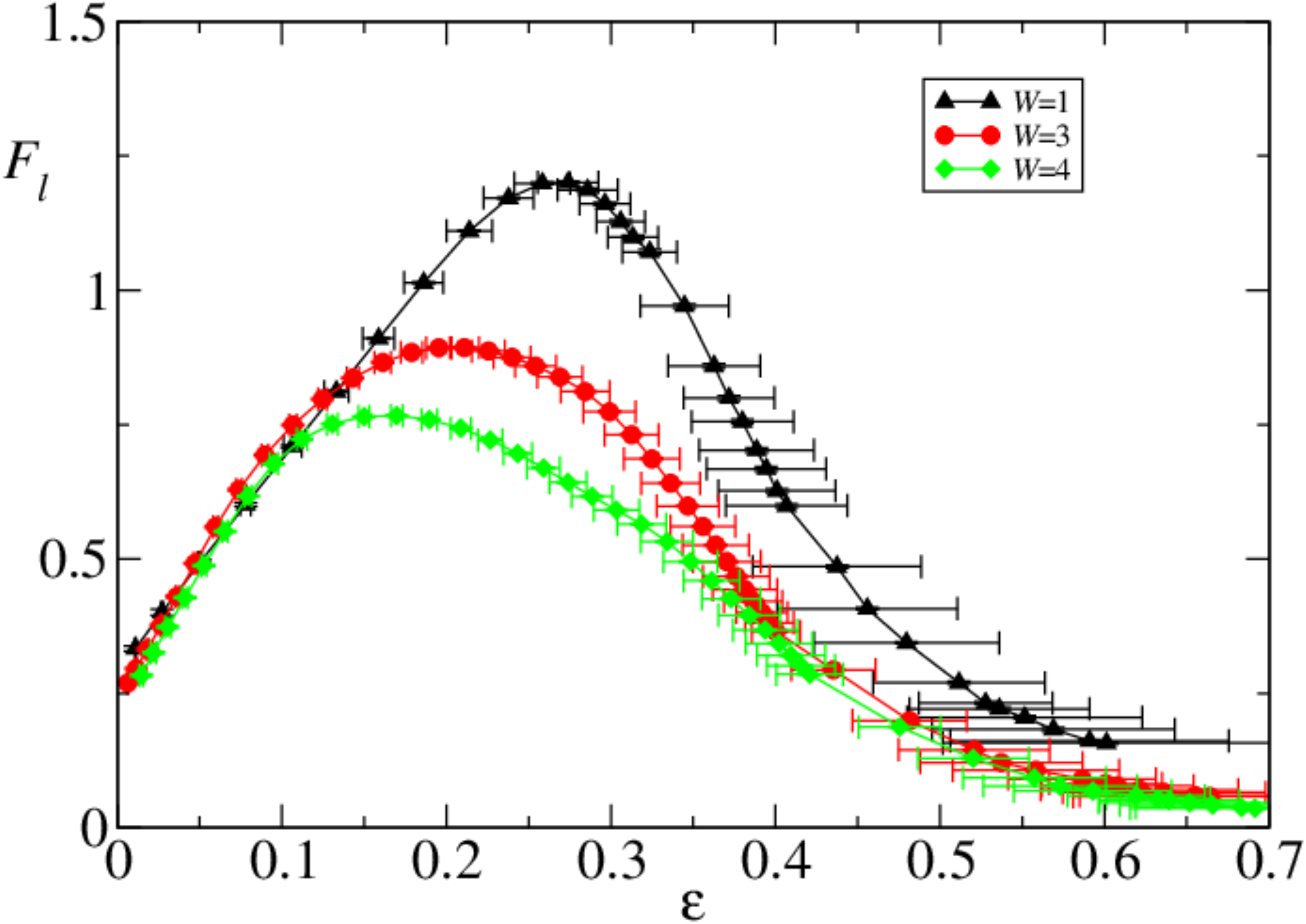}
  \caption{ {Long time mean edge fluctuation $F_l$ reveals the mobility edge and does not require 2-sites correlation function. The resolution is only slightly worse than that obtained
  with the mean fluctuation velocity $V_l$ as plotted in Fig.~5 of the main text. Data for L=8.
    \label{flsup}
 }}
\end{figure}
This is further visualized in Fig.~\ref{flsup} where mean site fluctuation (averaged over $t\in[220,250]$ in the similar manner to that for $\Delta x_L$), $F_l$ is plotted for different disorder values $W$. Note that already this observable confirms the sensitivity of the crossover to the excitation energy at different disorder values. Of course, similarly to the edge-fidelity derivative evaluation of fidelity does not require 2-point correlation function and involves local measurement only so it may be a method of choice for larger systems. Importantly
the fluctuations in the center of the chain are much less sensitive and do not reveal any transition in a consistent way.


\begin{thebibliography}{43}%
\makeatletter
\providecommand \@ifxundefined [1]{%
 \@ifx{#1\undefined}
}%
\providecommand \@ifnum [1]{%
 \ifnum #1\expandafter \@firstoftwo
 \else \expandafter \@secondoftwo
 \fi
}%
\providecommand \@ifx [1]{%
 \ifx #1\expandafter \@firstoftwo
 \else \expandafter \@secondoftwo
 \fi
}%
\providecommand \natexlab [1]{#1}%
\providecommand \enquote  [1]{``#1''}%
\providecommand \bibnamefont  [1]{#1}%
\providecommand \bibfnamefont [1]{#1}%
\providecommand \citenamefont [1]{#1}%
\providecommand \href@noop [0]{\@secondoftwo}%
\providecommand \href [0]{\begingroup \@sanitize@url \@href}%
\providecommand \@href[1]{\@@startlink{#1}\@@href}%
\providecommand \@@href[1]{\endgroup#1\@@endlink}%
\providecommand \@sanitize@url [0]{\catcode `\\12\catcode `\$12\catcode
  `\&12\catcode `\#12\catcode `\^12\catcode `\_12\catcode `\%12\relax}%
\providecommand \@@startlink[1]{}%
\providecommand \@@endlink[0]{}%
\providecommand \url  [0]{\begingroup\@sanitize@url \@url }%
\providecommand \@url [1]{\endgroup\@href {#1}{\urlprefix }}%
\providecommand \urlprefix  [0]{URL }%
\providecommand \Eprint [0]{\href }%
\providecommand \doibase [0]{http://dx.doi.org/}%
\providecommand \selectlanguage [0]{\@gobble}%
\providecommand \bibinfo  [0]{\@secondoftwo}%
\providecommand \bibfield  [0]{\@secondoftwo}%
\providecommand \translation [1]{[#1]}%
\providecommand \BibitemOpen [0]{}%
\providecommand \bibitemStop [0]{}%
\providecommand \bibitemNoStop [0]{.\EOS\space}%
\providecommand \EOS [0]{\spacefactor3000\relax}%
\providecommand \BibitemShut  [1]{\csname bibitem#1\endcsname}%
\let\auto@bib@innerbib\@empty
\bibitem [{\citenamefont {Huse}\ \emph {et~al.}(2014)\citenamefont {Huse},
  \citenamefont {Nandkishore},\ and\ \citenamefont {Oganesyan}}]{Huse14}%
  \BibitemOpen
  \bibfield  {author} {\bibinfo {author} {\bibfnamefont {D.~A.}\ \bibnamefont
  {Huse}}, \bibinfo {author} {\bibfnamefont {R.}~\bibnamefont {Nandkishore}}, \
  and\ \bibinfo {author} {\bibfnamefont {V.}~\bibnamefont {Oganesyan}},\ }\href
  {\doibase 10.1103/PhysRevB.90.174202} {\bibfield  {journal} {\bibinfo
  {journal} {Phys. Rev. B}\ }\textbf {\bibinfo {volume} {90}},\ \bibinfo
  {pages} {174202} (\bibinfo {year} {2014})}\BibitemShut {NoStop}%
\bibitem [{\citenamefont {Nandkishore}\ and\ \citenamefont
  {Huse}(2015)}]{Nandkishore15}%
  \BibitemOpen
  \bibfield  {author} {\bibinfo {author} {\bibfnamefont {R.}~\bibnamefont
  {Nandkishore}}\ and\ \bibinfo {author} {\bibfnamefont {D.~A.}\ \bibnamefont
  {Huse}},\ }\href@noop {} {\bibfield  {journal} {\bibinfo  {journal} {Ann.
  Rev. Cond. Mat. Phys.}\ }\textbf {\bibinfo {volume} {6}},\ \bibinfo {pages}
  {15} (\bibinfo {year} {2015})}\BibitemShut {NoStop}%
\bibitem [{\citenamefont {Alet}\ and\ \citenamefont
  {Laflorencie}(2018)}]{Alet18}%
  \BibitemOpen
  \bibfield  {author} {\bibinfo {author} {\bibfnamefont {F.}~\bibnamefont
  {Alet}}\ and\ \bibinfo {author} {\bibfnamefont {N.}~\bibnamefont
  {Laflorencie}},\ }\href {\doibase https://doi.org/10.1016/j.crhy.2018.03.003}
  {\bibfield  {journal} {\bibinfo  {journal} {Comptes Rendus Physique}\
  }\textbf {\bibinfo {volume} {19}},\ \bibinfo {pages} {498 } (\bibinfo {year}
  {2018})},\ \bibinfo {note} {quantum simulation / Simulation
  quantique}\BibitemShut {NoStop}%
\bibitem [{\citenamefont {Abanin}\ \emph
  {et~al.}(2019{\natexlab{a}})\citenamefont {Abanin}, \citenamefont {Altman},
  \citenamefont {Bloch},\ and\ \citenamefont {Serbyn}}]{Abanin19}%
  \BibitemOpen
  \bibfield  {author} {\bibinfo {author} {\bibfnamefont {D.~A.}\ \bibnamefont
  {Abanin}}, \bibinfo {author} {\bibfnamefont {E.}~\bibnamefont {Altman}},
  \bibinfo {author} {\bibfnamefont {I.}~\bibnamefont {Bloch}}, \ and\ \bibinfo
  {author} {\bibfnamefont {M.}~\bibnamefont {Serbyn}},\ }\href {\doibase
  10.1103/RevModPhys.91.021001} {\bibfield  {journal} {\bibinfo  {journal}
  {Rev. Mod. Phys.}\ }\textbf {\bibinfo {volume} {91}},\ \bibinfo {pages}
  {021001} (\bibinfo {year} {2019}{\natexlab{a}})}\BibitemShut {NoStop}%
\bibitem [{\citenamefont {{{\v{S}}untajs}}\ \emph {et~al.}(2019)\citenamefont
  {{{\v{S}}untajs}}, \citenamefont {{Bon{\v{c}}a}}, \citenamefont {{Prosen}},\
  and\ \citenamefont {{Vidmar}}}]{Suntais19}%
  \BibitemOpen
  \bibfield  {author} {\bibinfo {author} {\bibfnamefont {J.}~\bibnamefont
  {{{\v{S}}untajs}}}, \bibinfo {author} {\bibfnamefont {J.}~\bibnamefont
  {{Bon{\v{c}}a}}}, \bibinfo {author} {\bibfnamefont {T.}~\bibnamefont
  {{Prosen}}}, \ and\ \bibinfo {author} {\bibfnamefont {L.}~\bibnamefont
  {{Vidmar}}},\ }\href@noop {} {\bibfield  {journal} {\bibinfo  {journal}
  {arXiv e-prints}\ ,\ \bibinfo {pages} {arXiv:1905.06345}} (\bibinfo {year}
  {2019})},\ \Eprint {http://arxiv.org/abs/1905.06345} {arXiv:1905.06345
  [cond-mat.str-el]} \BibitemShut {NoStop}%
\bibitem [{\citenamefont {Abanin}\ \emph
  {et~al.}(2019{\natexlab{b}})\citenamefont {Abanin}, \citenamefont
  {Bardarson}, \citenamefont {Tomasi}, \citenamefont {Gopalakrishnan},
  \citenamefont {Khemani}, \citenamefont {Parameswaran}, \citenamefont
  {Pollmann}, \citenamefont {Potter}, \citenamefont {Serbyn},\ and\
  \citenamefont {Vasseur}}]{Abanin19z}%
  \BibitemOpen
  \bibfield  {author} {\bibinfo {author} {\bibfnamefont {D.~A.}\ \bibnamefont
  {Abanin}}, \bibinfo {author} {\bibfnamefont {J.~H.}\ \bibnamefont
  {Bardarson}}, \bibinfo {author} {\bibfnamefont {G.~D.}\ \bibnamefont
  {Tomasi}}, \bibinfo {author} {\bibfnamefont {S.}~\bibnamefont
  {Gopalakrishnan}}, \bibinfo {author} {\bibfnamefont {V.}~\bibnamefont
  {Khemani}}, \bibinfo {author} {\bibfnamefont {S.~A.}\ \bibnamefont
  {Parameswaran}}, \bibinfo {author} {\bibfnamefont {F.}~\bibnamefont
  {Pollmann}}, \bibinfo {author} {\bibfnamefont {A.~C.}\ \bibnamefont
  {Potter}}, \bibinfo {author} {\bibfnamefont {M.}~\bibnamefont {Serbyn}}, \
  and\ \bibinfo {author} {\bibfnamefont {R.}~\bibnamefont {Vasseur}},\
  }\href@noop {} {} (\bibinfo {year} {2019}{\natexlab{b}}),\ \Eprint
  {http://arxiv.org/abs/1911.04501} {arXiv:1911.04501 [cond-mat.str-el]}
  \BibitemShut {NoStop}%
\bibitem [{\citenamefont {Sierant}\ \emph {et~al.}(2019)\citenamefont
  {Sierant}, \citenamefont {Delande},\ and\ \citenamefont
  {Zakrzewski}}]{Sierant19z}%
  \BibitemOpen
  \bibfield  {author} {\bibinfo {author} {\bibfnamefont {P.}~\bibnamefont
  {Sierant}}, \bibinfo {author} {\bibfnamefont {D.}~\bibnamefont {Delande}}, \
  and\ \bibinfo {author} {\bibfnamefont {J.}~\bibnamefont {Zakrzewski}},\
  }\href@noop {} {} (\bibinfo {year} {2019}),\ \Eprint
  {http://arxiv.org/abs/1911.06221} {arXiv:1911.06221 [cond-mat.dis-nn]}
  \BibitemShut {NoStop}%
\bibitem [{\citenamefont {Panda}\ \emph {et~al.}(2019)\citenamefont {Panda},
  \citenamefont {Scardicchio}, \citenamefont {Schulz}, \citenamefont {Taylor},\
  and\ \citenamefont {Žnidarič}}]{Panda19}%
  \BibitemOpen
  \bibfield  {author} {\bibinfo {author} {\bibfnamefont {R.~K.}\ \bibnamefont
  {Panda}}, \bibinfo {author} {\bibfnamefont {A.}~\bibnamefont {Scardicchio}},
  \bibinfo {author} {\bibfnamefont {M.}~\bibnamefont {Schulz}}, \bibinfo
  {author} {\bibfnamefont {S.~R.}\ \bibnamefont {Taylor}}, \ and\ \bibinfo
  {author} {\bibfnamefont {M.}~\bibnamefont {Žnidarič}},\ }\href@noop {} {}
  (\bibinfo {year} {2019}),\ \Eprint {http://arxiv.org/abs/1911.07882}
  {arXiv:1911.07882 [cond-mat.dis-nn]} \BibitemShut {NoStop}%
\bibitem [{\citenamefont {Doggen}\ \emph {et~al.}(2018)\citenamefont {Doggen},
  \citenamefont {Schindler}, \citenamefont {Tikhonov}, \citenamefont {Mirlin},
  \citenamefont {Neupert}, \citenamefont {Polyakov},\ and\ \citenamefont
  {Gornyi}}]{Doggen18}%
  \BibitemOpen
  \bibfield  {author} {\bibinfo {author} {\bibfnamefont {E.~V.~H.}\
  \bibnamefont {Doggen}}, \bibinfo {author} {\bibfnamefont {F.}~\bibnamefont
  {Schindler}}, \bibinfo {author} {\bibfnamefont {K.~S.}\ \bibnamefont
  {Tikhonov}}, \bibinfo {author} {\bibfnamefont {A.~D.}\ \bibnamefont
  {Mirlin}}, \bibinfo {author} {\bibfnamefont {T.}~\bibnamefont {Neupert}},
  \bibinfo {author} {\bibfnamefont {D.~G.}\ \bibnamefont {Polyakov}}, \ and\
  \bibinfo {author} {\bibfnamefont {I.~V.}\ \bibnamefont {Gornyi}},\ }\href
  {\doibase 10.1103/PhysRevB.98.174202} {\bibfield  {journal} {\bibinfo
  {journal} {Phys. Rev. B}\ }\textbf {\bibinfo {volume} {98}},\ \bibinfo
  {pages} {174202} (\bibinfo {year} {2018})}\BibitemShut {NoStop}%
\bibitem [{\citenamefont {Chanda}\ \emph {et~al.}(2020)\citenamefont {Chanda},
  \citenamefont {Sierant},\ and\ \citenamefont {Zakrzewski}}]{Chanda19}%
  \BibitemOpen
  \bibfield  {author} {\bibinfo {author} {\bibfnamefont {T.}~\bibnamefont
  {Chanda}}, \bibinfo {author} {\bibfnamefont {P.}~\bibnamefont {Sierant}}, \
  and\ \bibinfo {author} {\bibfnamefont {J.}~\bibnamefont {Zakrzewski}},\
  }\href {\doibase 10.1103/PhysRevB.101.035148} {\bibfield  {journal} {\bibinfo
   {journal} {Phys. Rev. B}\ }\textbf {\bibinfo {volume} {101}},\ \bibinfo
  {pages} {035148} (\bibinfo {year} {2020})}\BibitemShut {NoStop}%
\bibitem [{\citenamefont {{Altman}}\ \emph {et~al.}(2019)\citenamefont
  {{Altman}}, \citenamefont {{Brown}}, \citenamefont {{Carleo}}, \citenamefont
  {{Carr}}, \citenamefont {{Demler}}, \citenamefont {{Chin}}, \citenamefont
  {{DeMarco}}, \citenamefont {{Economou}}, \citenamefont {{Eriksson}},
  \citenamefont {{Fu}}, \citenamefont {{Greiner}}, \citenamefont {{Hazzard}},
  \citenamefont {{Hulet}}, \citenamefont {{Kollar}}, \citenamefont {{Lev}},
  \citenamefont {{Lukin}}, \citenamefont {{Ma}}, \citenamefont {{Mi}},
  \citenamefont {{Misra}}, \citenamefont {{Monroe}}, \citenamefont {{Murch}},
  \citenamefont {{Nazario}}, \citenamefont {{Ni}}, \citenamefont {{Potter}},
  \citenamefont {{Roushan}}, \citenamefont {{Saffman}}, \citenamefont
  {{Schleier-Smith}}, \citenamefont {{Siddiqi}}, \citenamefont {{Simmonds}},
  \citenamefont {{Singh}}, \citenamefont {{Spielman}}, \citenamefont {{Temme}},
  \citenamefont {{Weiss}}, \citenamefont {{Vuckovic}}, \citenamefont
  {{Vuletic}}, \citenamefont {{Ye}},\ and\ \citenamefont
  {{Zwierlein}}}]{Altman19sim}%
  \BibitemOpen
  \bibfield  {author} {\bibinfo {author} {\bibfnamefont {E.}~\bibnamefont
  {{Altman}}}, \bibinfo {author} {\bibfnamefont {K.~R.}\ \bibnamefont
  {{Brown}}}, \bibinfo {author} {\bibfnamefont {G.}~\bibnamefont {{Carleo}}},
  \bibinfo {author} {\bibfnamefont {L.~D.}\ \bibnamefont {{Carr}}}, \bibinfo
  {author} {\bibfnamefont {E.}~\bibnamefont {{Demler}}}, \bibinfo {author}
  {\bibfnamefont {C.}~\bibnamefont {{Chin}}}, \bibinfo {author} {\bibfnamefont
  {B.}~\bibnamefont {{DeMarco}}}, \bibinfo {author} {\bibfnamefont {S.~E.}\
  \bibnamefont {{Economou}}}, \bibinfo {author} {\bibfnamefont {M.~A.}\
  \bibnamefont {{Eriksson}}}, \bibinfo {author} {\bibfnamefont {K.-M.~C.}\
  \bibnamefont {{Fu}}}, \bibinfo {author} {\bibfnamefont {M.}~\bibnamefont
  {{Greiner}}}, \bibinfo {author} {\bibfnamefont {K.~R.~A.}\ \bibnamefont
  {{Hazzard}}}, \bibinfo {author} {\bibfnamefont {R.~G.}\ \bibnamefont
  {{Hulet}}}, \bibinfo {author} {\bibfnamefont {A.~J.}\ \bibnamefont
  {{Kollar}}}, \bibinfo {author} {\bibfnamefont {B.~L.}\ \bibnamefont {{Lev}}},
  \bibinfo {author} {\bibfnamefont {M.~D.}\ \bibnamefont {{Lukin}}}, \bibinfo
  {author} {\bibfnamefont {R.}~\bibnamefont {{Ma}}}, \bibinfo {author}
  {\bibfnamefont {X.}~\bibnamefont {{Mi}}}, \bibinfo {author} {\bibfnamefont
  {S.}~\bibnamefont {{Misra}}}, \bibinfo {author} {\bibfnamefont
  {C.}~\bibnamefont {{Monroe}}}, \bibinfo {author} {\bibfnamefont
  {K.}~\bibnamefont {{Murch}}}, \bibinfo {author} {\bibfnamefont
  {Z.}~\bibnamefont {{Nazario}}}, \bibinfo {author} {\bibfnamefont {K.-K.}\
  \bibnamefont {{Ni}}}, \bibinfo {author} {\bibfnamefont {A.~C.}\ \bibnamefont
  {{Potter}}}, \bibinfo {author} {\bibfnamefont {P.}~\bibnamefont {{Roushan}}},
  \bibinfo {author} {\bibfnamefont {M.}~\bibnamefont {{Saffman}}}, \bibinfo
  {author} {\bibfnamefont {M.}~\bibnamefont {{Schleier-Smith}}}, \bibinfo
  {author} {\bibfnamefont {I.}~\bibnamefont {{Siddiqi}}}, \bibinfo {author}
  {\bibfnamefont {R.}~\bibnamefont {{Simmonds}}}, \bibinfo {author}
  {\bibfnamefont {M.}~\bibnamefont {{Singh}}}, \bibinfo {author} {\bibfnamefont
  {I.~B.}\ \bibnamefont {{Spielman}}}, \bibinfo {author} {\bibfnamefont
  {K.}~\bibnamefont {{Temme}}}, \bibinfo {author} {\bibfnamefont {D.~S.}\
  \bibnamefont {{Weiss}}}, \bibinfo {author} {\bibfnamefont {J.}~\bibnamefont
  {{Vuckovic}}}, \bibinfo {author} {\bibfnamefont {V.}~\bibnamefont
  {{Vuletic}}}, \bibinfo {author} {\bibfnamefont {J.}~\bibnamefont {{Ye}}}, \
  and\ \bibinfo {author} {\bibfnamefont {M.}~\bibnamefont {{Zwierlein}}},\
  }\href@noop {} {\bibfield  {journal} {\bibinfo  {journal} {arXiv e-prints}\
  ,\ \bibinfo {eid} {arXiv:1912.06938}} (\bibinfo {year} {2019})},\ \Eprint
  {http://arxiv.org/abs/1912.06938} {arXiv:1912.06938 [quant-ph]} \BibitemShut
  {NoStop}%
\bibitem [{\citenamefont {Kondov}\ \emph {et~al.}(2015)\citenamefont {Kondov},
  \citenamefont {McGehee}, \citenamefont {Xu},\ and\ \citenamefont
  {DeMarco}}]{Kondov14}%
  \BibitemOpen
  \bibfield  {author} {\bibinfo {author} {\bibfnamefont {S.~S.}\ \bibnamefont
  {Kondov}}, \bibinfo {author} {\bibfnamefont {W.~R.}\ \bibnamefont {McGehee}},
  \bibinfo {author} {\bibfnamefont {W.}~\bibnamefont {Xu}}, \ and\ \bibinfo
  {author} {\bibfnamefont {B.}~\bibnamefont {DeMarco}},\ }\href {\doibase
  10.1103/PhysRevLett.114.083002} {\bibfield  {journal} {\bibinfo  {journal}
  {Phys. Rev. Lett.}\ }\textbf {\bibinfo {volume} {114}},\ \bibinfo {pages}
  {083002} (\bibinfo {year} {2015})}\BibitemShut {NoStop}%
\bibitem [{\citenamefont {Schreiber}\ \emph {et~al.}(2015)\citenamefont
  {Schreiber}, \citenamefont {Hodgman}, \citenamefont {Bordia}, \citenamefont
  {L{\"u}schen}, \citenamefont {Fischer}, \citenamefont {Vosk}, \citenamefont
  {Altman}, \citenamefont {Schneider},\ and\ \citenamefont
  {Bloch}}]{Schreiber15}%
  \BibitemOpen
  \bibfield  {author} {\bibinfo {author} {\bibfnamefont {M.}~\bibnamefont
  {Schreiber}}, \bibinfo {author} {\bibfnamefont {S.~S.}\ \bibnamefont
  {Hodgman}}, \bibinfo {author} {\bibfnamefont {P.}~\bibnamefont {Bordia}},
  \bibinfo {author} {\bibfnamefont {H.~P.}\ \bibnamefont {L{\"u}schen}},
  \bibinfo {author} {\bibfnamefont {M.~H.}\ \bibnamefont {Fischer}}, \bibinfo
  {author} {\bibfnamefont {R.}~\bibnamefont {Vosk}}, \bibinfo {author}
  {\bibfnamefont {E.}~\bibnamefont {Altman}}, \bibinfo {author} {\bibfnamefont
  {U.}~\bibnamefont {Schneider}}, \ and\ \bibinfo {author} {\bibfnamefont
  {I.}~\bibnamefont {Bloch}},\ }\href {\doibase 10.1126/science.aaa7432}
  {\bibfield  {journal} {\bibinfo  {journal} {Science}\ }\textbf {\bibinfo
  {volume} {349}},\ \bibinfo {pages} {842} (\bibinfo {year}
  {2015})}\BibitemShut {NoStop}%
\bibitem [{\citenamefont {L\"uschen}\ \emph {et~al.}(2017)\citenamefont
  {L\"uschen}, \citenamefont {Bordia}, \citenamefont {Scherg}, \citenamefont
  {Alet}, \citenamefont {Altman}, \citenamefont {Schneider},\ and\
  \citenamefont {Bloch}}]{Luschen17}%
  \BibitemOpen
  \bibfield  {author} {\bibinfo {author} {\bibfnamefont {H.~P.}\ \bibnamefont
  {L\"uschen}}, \bibinfo {author} {\bibfnamefont {P.}~\bibnamefont {Bordia}},
  \bibinfo {author} {\bibfnamefont {S.}~\bibnamefont {Scherg}}, \bibinfo
  {author} {\bibfnamefont {F.}~\bibnamefont {Alet}}, \bibinfo {author}
  {\bibfnamefont {E.}~\bibnamefont {Altman}}, \bibinfo {author} {\bibfnamefont
  {U.}~\bibnamefont {Schneider}}, \ and\ \bibinfo {author} {\bibfnamefont
  {I.}~\bibnamefont {Bloch}},\ }\href {\doibase 10.1103/PhysRevLett.119.260401}
  {\bibfield  {journal} {\bibinfo  {journal} {Phys. Rev. Lett.}\ }\textbf
  {\bibinfo {volume} {119}},\ \bibinfo {pages} {260401} (\bibinfo {year}
  {2017})}\BibitemShut {NoStop}%
\bibitem [{\citenamefont {Choi}\ \emph {et~al.}(2016)\citenamefont {Choi},
  \citenamefont {Hild}, \citenamefont {Zeiher}, \citenamefont {Schau{\ss}},
  \citenamefont {Rubio-Abadal}, \citenamefont {Yefsah}, \citenamefont
  {Khemani}, \citenamefont {Huse}, \citenamefont {Bloch},\ and\ \citenamefont
  {Gross}}]{Choi16}%
  \BibitemOpen
  \bibfield  {author} {\bibinfo {author} {\bibfnamefont {J.-y.}\ \bibnamefont
  {Choi}}, \bibinfo {author} {\bibfnamefont {S.}~\bibnamefont {Hild}}, \bibinfo
  {author} {\bibfnamefont {J.}~\bibnamefont {Zeiher}}, \bibinfo {author}
  {\bibfnamefont {P.}~\bibnamefont {Schau{\ss}}}, \bibinfo {author}
  {\bibfnamefont {A.}~\bibnamefont {Rubio-Abadal}}, \bibinfo {author}
  {\bibfnamefont {T.}~\bibnamefont {Yefsah}}, \bibinfo {author} {\bibfnamefont
  {V.}~\bibnamefont {Khemani}}, \bibinfo {author} {\bibfnamefont {D.~A.}\
  \bibnamefont {Huse}}, \bibinfo {author} {\bibfnamefont {I.}~\bibnamefont
  {Bloch}}, \ and\ \bibinfo {author} {\bibfnamefont {C.}~\bibnamefont
  {Gross}},\ }\href {\doibase 10.1126/science.aaf8834} {\bibfield  {journal}
  {\bibinfo  {journal} {Science}\ }\textbf {\bibinfo {volume} {352}},\ \bibinfo
  {pages} {1547} (\bibinfo {year} {2016})}\BibitemShut {NoStop}%
\bibitem [{\citenamefont {Roushan}\ \emph {et~al.}(2017)\citenamefont
  {Roushan}, \citenamefont {Neill}, \citenamefont {Tangpanitanon},
  \citenamefont {Bastidas}, \citenamefont {Megrant}, \citenamefont {Barends},
  \citenamefont {Chen}, \citenamefont {Chen}, \citenamefont {Chiaro},
  \citenamefont {Dunsworth}, \citenamefont {Fowler}, \citenamefont {Foxen},
  \citenamefont {Giustina}, \citenamefont {Jeffrey}, \citenamefont {Kelly},
  \citenamefont {Lucero}, \citenamefont {Mutus}, \citenamefont {Neeley},
  \citenamefont {Quintana}, \citenamefont {Sank}, \citenamefont {Vainsencher},
  \citenamefont {Wenner}, \citenamefont {White}, \citenamefont {Neven},
  \citenamefont {Angelakis},\ and\ \citenamefont {Martinis}}]{Roushan17}%
  \BibitemOpen
  \bibfield  {author} {\bibinfo {author} {\bibfnamefont {P.}~\bibnamefont
  {Roushan}}, \bibinfo {author} {\bibfnamefont {C.}~\bibnamefont {Neill}},
  \bibinfo {author} {\bibfnamefont {J.}~\bibnamefont {Tangpanitanon}}, \bibinfo
  {author} {\bibfnamefont {V.~M.}\ \bibnamefont {Bastidas}}, \bibinfo {author}
  {\bibfnamefont {A.}~\bibnamefont {Megrant}}, \bibinfo {author} {\bibfnamefont
  {R.}~\bibnamefont {Barends}}, \bibinfo {author} {\bibfnamefont
  {Y.}~\bibnamefont {Chen}}, \bibinfo {author} {\bibfnamefont {Z.}~\bibnamefont
  {Chen}}, \bibinfo {author} {\bibfnamefont {B.}~\bibnamefont {Chiaro}},
  \bibinfo {author} {\bibfnamefont {A.}~\bibnamefont {Dunsworth}}, \bibinfo
  {author} {\bibfnamefont {A.}~\bibnamefont {Fowler}}, \bibinfo {author}
  {\bibfnamefont {B.}~\bibnamefont {Foxen}}, \bibinfo {author} {\bibfnamefont
  {M.}~\bibnamefont {Giustina}}, \bibinfo {author} {\bibfnamefont
  {E.}~\bibnamefont {Jeffrey}}, \bibinfo {author} {\bibfnamefont
  {J.}~\bibnamefont {Kelly}}, \bibinfo {author} {\bibfnamefont
  {E.}~\bibnamefont {Lucero}}, \bibinfo {author} {\bibfnamefont
  {J.}~\bibnamefont {Mutus}}, \bibinfo {author} {\bibfnamefont
  {M.}~\bibnamefont {Neeley}}, \bibinfo {author} {\bibfnamefont
  {C.}~\bibnamefont {Quintana}}, \bibinfo {author} {\bibfnamefont
  {D.}~\bibnamefont {Sank}}, \bibinfo {author} {\bibfnamefont {A.}~\bibnamefont
  {Vainsencher}}, \bibinfo {author} {\bibfnamefont {J.}~\bibnamefont {Wenner}},
  \bibinfo {author} {\bibfnamefont {T.}~\bibnamefont {White}}, \bibinfo
  {author} {\bibfnamefont {H.}~\bibnamefont {Neven}}, \bibinfo {author}
  {\bibfnamefont {D.~G.}\ \bibnamefont {Angelakis}}, \ and\ \bibinfo {author}
  {\bibfnamefont {J.}~\bibnamefont {Martinis}},\ }\href {\doibase
  10.1126/science.aao1401} {\bibfield  {journal} {\bibinfo  {journal}
  {Science}\ }\textbf {\bibinfo {volume} {358}},\ \bibinfo {pages} {1175}
  (\bibinfo {year} {2017})},\ \Eprint
  {http://arxiv.org/abs/https://science.sciencemag.org/content/358/6367/1175.full.pdf}
  {https://science.sciencemag.org/content/358/6367/1175.full.pdf} \BibitemShut
  {NoStop}%
\bibitem [{\citenamefont {Lukin}\ \emph {et~al.}(2019)\citenamefont {Lukin},
  \citenamefont {Rispoli}, \citenamefont {Schittko}, \citenamefont {Tai},
  \citenamefont {Kaufman}, \citenamefont {Choi}, \citenamefont {Khemani},
  \citenamefont {L{\'e}onard},\ and\ \citenamefont {Greiner}}]{Lukin19}%
  \BibitemOpen
  \bibfield  {author} {\bibinfo {author} {\bibfnamefont {A.}~\bibnamefont
  {Lukin}}, \bibinfo {author} {\bibfnamefont {M.}~\bibnamefont {Rispoli}},
  \bibinfo {author} {\bibfnamefont {R.}~\bibnamefont {Schittko}}, \bibinfo
  {author} {\bibfnamefont {M.~E.}\ \bibnamefont {Tai}}, \bibinfo {author}
  {\bibfnamefont {A.~M.}\ \bibnamefont {Kaufman}}, \bibinfo {author}
  {\bibfnamefont {S.}~\bibnamefont {Choi}}, \bibinfo {author} {\bibfnamefont
  {V.}~\bibnamefont {Khemani}}, \bibinfo {author} {\bibfnamefont
  {J.}~\bibnamefont {L{\'e}onard}}, \ and\ \bibinfo {author} {\bibfnamefont
  {M.}~\bibnamefont {Greiner}},\ }\href {\doibase 10.1126/science.aau0818}
  {\bibfield  {journal} {\bibinfo  {journal} {Science}\ }\textbf {\bibinfo
  {volume} {364}},\ \bibinfo {pages} {256} (\bibinfo {year} {2019})},\ \Eprint
  {http://arxiv.org/abs/https://science.sciencemag.org/content/364/6437/256.full.pdf}
  {https://science.sciencemag.org/content/364/6437/256.full.pdf} \BibitemShut
  {NoStop}%
\bibitem [{\citenamefont {Rispoli}\ \emph {et~al.}(2019)\citenamefont
  {Rispoli}, \citenamefont {Lukin}, \citenamefont {Schittko}, \citenamefont
  {Kim}, \citenamefont {Tai}, \citenamefont {L{\'e}onard},\ and\ \citenamefont
  {Greiner}}]{Rispoli19}%
  \BibitemOpen
  \bibfield  {author} {\bibinfo {author} {\bibfnamefont {M.}~\bibnamefont
  {Rispoli}}, \bibinfo {author} {\bibfnamefont {A.}~\bibnamefont {Lukin}},
  \bibinfo {author} {\bibfnamefont {R.}~\bibnamefont {Schittko}}, \bibinfo
  {author} {\bibfnamefont {S.}~\bibnamefont {Kim}}, \bibinfo {author}
  {\bibfnamefont {M.~E.}\ \bibnamefont {Tai}}, \bibinfo {author} {\bibfnamefont
  {J.}~\bibnamefont {L{\'e}onard}}, \ and\ \bibinfo {author} {\bibfnamefont
  {M.}~\bibnamefont {Greiner}},\ }\href {\doibase 10.1038/s41586-019-1527-2}
  {\bibfield  {journal} {\bibinfo  {journal} {Nature}\ }\textbf {\bibinfo
  {volume} {573}},\ \bibinfo {pages} {385} (\bibinfo {year}
  {2019})}\BibitemShut {NoStop}%
\bibitem [{\citenamefont {Wahl}\ \emph {et~al.}(2019)\citenamefont {Wahl},
  \citenamefont {Pal},\ and\ \citenamefont {Simon}}]{Wahl19}%
  \BibitemOpen
  \bibfield  {author} {\bibinfo {author} {\bibfnamefont {T.~B.}\ \bibnamefont
  {Wahl}}, \bibinfo {author} {\bibfnamefont {A.}~\bibnamefont {Pal}}, \ and\
  \bibinfo {author} {\bibfnamefont {S.~H.}\ \bibnamefont {Simon}},\ }\href
  {\doibase 10.1038/s41567-018-0339-x} {\bibfield  {journal} {\bibinfo
  {journal} {Nature Physics}\ }\textbf {\bibinfo {volume} {15}},\ \bibinfo
  {pages} {164} (\bibinfo {year} {2019})}\BibitemShut {NoStop}%
\bibitem [{\citenamefont {Sierant}\ and\ \citenamefont
  {Zakrzewski}(2018)}]{Sierant18}%
  \BibitemOpen
  \bibfield  {author} {\bibinfo {author} {\bibfnamefont {P.}~\bibnamefont
  {Sierant}}\ and\ \bibinfo {author} {\bibfnamefont {J.}~\bibnamefont
  {Zakrzewski}},\ }\href {\doibase 10.1088/1367-2630/aabb17} {\bibfield
  {journal} {\bibinfo  {journal} {New Journal of Physics}\ }\textbf {\bibinfo
  {volume} {20}},\ \bibinfo {pages} {043032} (\bibinfo {year}
  {2018})}\BibitemShut {NoStop}%
\bibitem [{\citenamefont {Orell}\ \emph {et~al.}(2019)\citenamefont {Orell},
  \citenamefont {Michailidis}, \citenamefont {Serbyn},\ and\ \citenamefont
  {Silveri}}]{Orel19}%
  \BibitemOpen
  \bibfield  {author} {\bibinfo {author} {\bibfnamefont {T.}~\bibnamefont
  {Orell}}, \bibinfo {author} {\bibfnamefont {A.~A.}\ \bibnamefont
  {Michailidis}}, \bibinfo {author} {\bibfnamefont {M.}~\bibnamefont {Serbyn}},
  \ and\ \bibinfo {author} {\bibfnamefont {M.}~\bibnamefont {Silveri}},\ }\href
  {\doibase 10.1103/PhysRevB.100.134504} {\bibfield  {journal} {\bibinfo
  {journal} {Phys. Rev. B}\ }\textbf {\bibinfo {volume} {100}},\ \bibinfo
  {pages} {134504} (\bibinfo {year} {2019})}\BibitemShut {NoStop}%
\bibitem [{\citenamefont {{Krause}}\ \emph {et~al.}(2019)\citenamefont
  {{Krause}}, \citenamefont {{Pellegrin}}, \citenamefont {{Brouwer}},
  \citenamefont {{Abanin}},\ and\ \citenamefont {{Filippone}}}]{Krause19}%
  \BibitemOpen
  \bibfield  {author} {\bibinfo {author} {\bibfnamefont {U.}~\bibnamefont
  {{Krause}}}, \bibinfo {author} {\bibfnamefont {T.}~\bibnamefont
  {{Pellegrin}}}, \bibinfo {author} {\bibfnamefont {P.~W.}\ \bibnamefont
  {{Brouwer}}}, \bibinfo {author} {\bibfnamefont {D.~A.}\ \bibnamefont
  {{Abanin}}}, \ and\ \bibinfo {author} {\bibfnamefont {M.}~\bibnamefont
  {{Filippone}}},\ }\href@noop {} {\bibfield  {journal} {\bibinfo  {journal}
  {arXiv e-prints}\ ,\ \bibinfo {eid} {arXiv:1911.11711}} (\bibinfo {year}
  {2019})},\ \Eprint {http://arxiv.org/abs/1911.11711} {arXiv:1911.11711
  [cond-mat.dis-nn]} \BibitemShut {NoStop}%
\bibitem [{\citenamefont {{Hopjan}}\ and\ \citenamefont
  {{Heidrich-Meisner}}(2019)}]{Hopjan19}%
  \BibitemOpen
  \bibfield  {author} {\bibinfo {author} {\bibfnamefont {M.}~\bibnamefont
  {{Hopjan}}}\ and\ \bibinfo {author} {\bibfnamefont {F.}~\bibnamefont
  {{Heidrich-Meisner}}},\ }\href@noop {} {\bibfield  {journal} {\bibinfo
  {journal} {arXiv e-prints}\ ,\ \bibinfo {eid} {arXiv:1912.09443}} (\bibinfo
  {year} {2019})},\ \Eprint {http://arxiv.org/abs/1912.09443} {arXiv:1912.09443
  [cond-mat.str-el]} \BibitemShut {NoStop}%
\bibitem [{\citenamefont {Sierant}\ \emph {et~al.}(2017)\citenamefont
  {Sierant}, \citenamefont {Delande},\ and\ \citenamefont
  {Zakrzewski}}]{Sierant17}%
  \BibitemOpen
  \bibfield  {author} {\bibinfo {author} {\bibfnamefont {P.}~\bibnamefont
  {Sierant}}, \bibinfo {author} {\bibfnamefont {D.}~\bibnamefont {Delande}}, \
  and\ \bibinfo {author} {\bibfnamefont {J.}~\bibnamefont {Zakrzewski}},\
  }\href {\doibase 10.1103/PhysRevA.95.021601} {\bibfield  {journal} {\bibinfo
  {journal} {Phys. Rev. A}\ }\textbf {\bibinfo {volume} {95}},\ \bibinfo
  {pages} {021601} (\bibinfo {year} {2017})}\BibitemShut {NoStop}%
\bibitem [{\citenamefont {{Sierant}}\ \emph {et~al.}(2017)\citenamefont
  {{Sierant}}, \citenamefont {{Delande}},\ and\ \citenamefont
  {{Zakrzewski}}}]{Sierant17b}%
  \BibitemOpen
  \bibfield  {author} {\bibinfo {author} {\bibfnamefont {P.}~\bibnamefont
  {{Sierant}}}, \bibinfo {author} {\bibfnamefont {D.}~\bibnamefont
  {{Delande}}}, \ and\ \bibinfo {author} {\bibfnamefont {J.}~\bibnamefont
  {{Zakrzewski}}},\ }\href {\doibase 10.12693/APhysPolA.132.1707} {\bibfield
  {journal} {\bibinfo  {journal} {Acta Phys. Polon. A}\ }\textbf {\bibinfo
  {volume} {132}},\ \bibinfo {pages} {1707} (\bibinfo {year}
  {2017})}\BibitemShut {NoStop}%
\bibitem [{\citenamefont {Guarrera}\ \emph {et~al.}(2007)\citenamefont
  {Guarrera}, \citenamefont {Fallani}, \citenamefont {Lye}, \citenamefont
  {Fort},\ and\ \citenamefont {Inguscio}}]{Guarrera07}%
  \BibitemOpen
  \bibfield  {author} {\bibinfo {author} {\bibfnamefont {V.}~\bibnamefont
  {Guarrera}}, \bibinfo {author} {\bibfnamefont {L.}~\bibnamefont {Fallani}},
  \bibinfo {author} {\bibfnamefont {J.~E.}\ \bibnamefont {Lye}}, \bibinfo
  {author} {\bibfnamefont {C.}~\bibnamefont {Fort}}, \ and\ \bibinfo {author}
  {\bibfnamefont {M.}~\bibnamefont {Inguscio}},\ }\href {\doibase
  10.1088/1367-2630/9/4/107} {\bibfield  {journal} {\bibinfo  {journal} {New J.
  Phys.}\ }\textbf {\bibinfo {volume} {9}},\ \bibinfo {pages} {107} (\bibinfo
  {year} {2007})}\BibitemShut {NoStop}%
\bibitem [{\citenamefont {Doggen}\ and\ \citenamefont
  {Mirlin}(2019)}]{Doggen19}%
  \BibitemOpen
  \bibfield  {author} {\bibinfo {author} {\bibfnamefont {E.~V.~H.}\
  \bibnamefont {Doggen}}\ and\ \bibinfo {author} {\bibfnamefont {A.~D.}\
  \bibnamefont {Mirlin}},\ }\href {\doibase 10.1103/PhysRevB.100.104203}
  {\bibfield  {journal} {\bibinfo  {journal} {Phys. Rev. B}\ }\textbf {\bibinfo
  {volume} {100}},\ \bibinfo {pages} {104203} (\bibinfo {year}
  {2019})}\BibitemShut {NoStop}%
\bibitem [{\citenamefont {Yao}\ \emph {et~al.}(2019)\citenamefont {Yao},
  \citenamefont {Khoudli}, \citenamefont {Bresque},\ and\ \citenamefont
  {Sanchez-Palencia}}]{Hepeng19}%
  \BibitemOpen
  \bibfield  {author} {\bibinfo {author} {\bibfnamefont {H.}~\bibnamefont
  {Yao}}, \bibinfo {author} {\bibfnamefont {H.}~\bibnamefont {Khoudli}},
  \bibinfo {author} {\bibfnamefont {L.}~\bibnamefont {Bresque}}, \ and\
  \bibinfo {author} {\bibfnamefont {L.}~\bibnamefont {Sanchez-Palencia}},\
  }\href {\doibase 10.1103/PhysRevLett.123.070405} {\bibfield  {journal}
  {\bibinfo  {journal} {Phys. Rev. Lett.}\ }\textbf {\bibinfo {volume} {123}},\
  \bibinfo {pages} {070405} (\bibinfo {year} {2019})}\BibitemShut {NoStop}%
\bibitem [{\citenamefont {Oganesyan}\ and\ \citenamefont
  {Huse}(2007)}]{Oganesyan07}%
  \BibitemOpen
  \bibfield  {author} {\bibinfo {author} {\bibfnamefont {V.}~\bibnamefont
  {Oganesyan}}\ and\ \bibinfo {author} {\bibfnamefont {D.~A.}\ \bibnamefont
  {Huse}},\ }\href {\doibase 10.1103/PhysRevB.75.155111} {\bibfield  {journal}
  {\bibinfo  {journal} {Phys. Rev. B}\ }\textbf {\bibinfo {volume} {75}},\
  \bibinfo {pages} {155111} (\bibinfo {year} {2007})}\BibitemShut {NoStop}%
\bibitem [{\citenamefont {Mehta}(1990)}]{Mehtabook}%
  \BibitemOpen
  \bibfield  {author} {\bibinfo {author} {\bibfnamefont {M.~L.}\ \bibnamefont
  {Mehta}},\ }\href@noop {} {\emph {\bibinfo {title} {Random Matrices}}}\
  (\bibinfo  {publisher} {Elsevier, Amsterdam},\ \bibinfo {year}
  {1990})\BibitemShut {NoStop}%
\bibitem [{\citenamefont {Haake}(2010)}]{Haake}%
  \BibitemOpen
  \bibfield  {author} {\bibinfo {author} {\bibfnamefont {F.}~\bibnamefont
  {Haake}},\ }\href@noop {} {\emph {\bibinfo {title} {Quantum Signatures of
  Chaos}}}\ (\bibinfo  {publisher} {Springer, Berlin},\ \bibinfo {year}
  {2010})\BibitemShut {NoStop}%
\bibitem [{\citenamefont {Luitz}\ \emph {et~al.}(2015)\citenamefont {Luitz},
  \citenamefont {Laflorencie},\ and\ \citenamefont {Alet}}]{Luitz15}%
  \BibitemOpen
  \bibfield  {author} {\bibinfo {author} {\bibfnamefont {D.~J.}\ \bibnamefont
  {Luitz}}, \bibinfo {author} {\bibfnamefont {N.}~\bibnamefont {Laflorencie}},
  \ and\ \bibinfo {author} {\bibfnamefont {F.}~\bibnamefont {Alet}},\ }\href
  {\doibase 10.1103/PhysRevB.91.081103} {\bibfield  {journal} {\bibinfo
  {journal} {Phys. Rev. B}\ }\textbf {\bibinfo {volume} {91}},\ \bibinfo
  {pages} {081103} (\bibinfo {year} {2015})}\BibitemShut {NoStop}%
\bibitem [{\citenamefont {Macé}\ \emph {et~al.}(2019)\citenamefont {Macé},
  \citenamefont {Laflorencie},\ and\ \citenamefont {Alet}}]{Mace19}%
  \BibitemOpen
  \bibfield  {author} {\bibinfo {author} {\bibfnamefont {N.}~\bibnamefont
  {Macé}}, \bibinfo {author} {\bibfnamefont {N.}~\bibnamefont {Laflorencie}},
  \ and\ \bibinfo {author} {\bibfnamefont {F.}~\bibnamefont {Alet}},\ }\href
  {\doibase 10.21468/SciPostPhys.6.4.050} {\bibfield  {journal} {\bibinfo
  {journal} {SciPost Phys.}\ }\textbf {\bibinfo {volume} {6}},\ \bibinfo
  {pages} {50} (\bibinfo {year} {2019})}\BibitemShut {NoStop}%
\bibitem [{\citenamefont {{Laflorencie}}\ \emph {et~al.}(2020)\citenamefont
  {{Laflorencie}}, \citenamefont {{Lemari{\'e}}},\ and\ \citenamefont
  {{Mac{\'e}}}}]{Laflorencie20}%
  \BibitemOpen
  \bibfield  {author} {\bibinfo {author} {\bibfnamefont {N.}~\bibnamefont
  {{Laflorencie}}}, \bibinfo {author} {\bibfnamefont {G.}~\bibnamefont
  {{Lemari{\'e}}}}, \ and\ \bibinfo {author} {\bibfnamefont {N.}~\bibnamefont
  {{Mac{\'e}}}},\ }\href@noop {} {\bibfield  {journal} {\bibinfo  {journal}
  {arXiv e-prints}\ ,\ \bibinfo {eid} {arXiv:2004.02861}} (\bibinfo {year}
  {2020})},\ \Eprint {http://arxiv.org/abs/2004.02861} {arXiv:2004.02861
  [cond-mat.dis-nn]} \BibitemShut {NoStop}%
\bibitem [{\citenamefont {{{\v{S}}untajs}}\ \emph {et~al.}(2020)\citenamefont
  {{{\v{S}}untajs}}, \citenamefont {{Bon{\v{c}}a}}, \citenamefont {{Prosen}},\
  and\ \citenamefont {{Vidmar}}}]{Suntajs20}%
  \BibitemOpen
  \bibfield  {author} {\bibinfo {author} {\bibfnamefont {J.}~\bibnamefont
  {{{\v{S}}untajs}}}, \bibinfo {author} {\bibfnamefont {J.}~\bibnamefont
  {{Bon{\v{c}}a}}}, \bibinfo {author} {\bibfnamefont {T.}~\bibnamefont
  {{Prosen}}}, \ and\ \bibinfo {author} {\bibfnamefont {L.}~\bibnamefont
  {{Vidmar}}},\ }\href@noop {} {\bibfield  {journal} {\bibinfo  {journal}
  {arXiv e-prints}\ ,\ \bibinfo {eid} {arXiv:2004.01719}} (\bibinfo {year}
  {2020})},\ \Eprint {http://arxiv.org/abs/2004.01719} {arXiv:2004.01719
  [cond-mat.dis-nn]} \BibitemShut {NoStop}%
\bibitem [{sup()}]{suppl}%
  \BibitemOpen
  \href@noop {} {}\bibinfo {note} {See Supplemental Material at [URL will be
  inserted by publisher] for additional details.}\BibitemShut {Stop}%
\bibitem [{\citenamefont {Bera}\ \emph {et~al.}(2017)\citenamefont {Bera},
  \citenamefont {De~Tomasi}, \citenamefont {Weiner},\ and\ \citenamefont
  {Evers}}]{Bera17}%
  \BibitemOpen
  \bibfield  {author} {\bibinfo {author} {\bibfnamefont {S.}~\bibnamefont
  {Bera}}, \bibinfo {author} {\bibfnamefont {G.}~\bibnamefont {De~Tomasi}},
  \bibinfo {author} {\bibfnamefont {F.}~\bibnamefont {Weiner}}, \ and\ \bibinfo
  {author} {\bibfnamefont {F.}~\bibnamefont {Evers}},\ }\href {\doibase
  10.1103/PhysRevLett.118.196801} {\bibfield  {journal} {\bibinfo  {journal}
  {Phys. Rev. Lett.}\ }\textbf {\bibinfo {volume} {118}},\ \bibinfo {pages}
  {196801} (\bibinfo {year} {2017})}\BibitemShut {NoStop}%
\bibitem [{\citenamefont {Weiner}\ \emph {et~al.}(2019)\citenamefont {Weiner},
  \citenamefont {Evers},\ and\ \citenamefont {Bera}}]{Weiner19}%
  \BibitemOpen
  \bibfield  {author} {\bibinfo {author} {\bibfnamefont {F.}~\bibnamefont
  {Weiner}}, \bibinfo {author} {\bibfnamefont {F.}~\bibnamefont {Evers}}, \
  and\ \bibinfo {author} {\bibfnamefont {S.}~\bibnamefont {Bera}},\ }\href
  {\doibase 10.1103/PhysRevB.100.104204} {\bibfield  {journal} {\bibinfo
  {journal} {Phys. Rev. B}\ }\textbf {\bibinfo {volume} {100}},\ \bibinfo
  {pages} {104204} (\bibinfo {year} {2019})}\BibitemShut {NoStop}%
\bibitem [{\citenamefont {Tal‐Ezer}\ and\ \citenamefont
  {Kosloff}(1984)}]{Kosloff84}%
  \BibitemOpen
  \bibfield  {author} {\bibinfo {author} {\bibfnamefont {H.}~\bibnamefont
  {Tal‐Ezer}}\ and\ \bibinfo {author} {\bibfnamefont {R.}~\bibnamefont
  {Kosloff}},\ }\href {\doibase 10.1063/1.448136} {\bibfield  {journal}
  {\bibinfo  {journal} {The Journal of Chemical Physics}\ }\textbf {\bibinfo
  {volume} {81}},\ \bibinfo {pages} {3967} (\bibinfo {year} {1984})},\ \Eprint
  {http://arxiv.org/abs/https://doi.org/10.1063/1.448136}
  {https://doi.org/10.1063/1.448136} \BibitemShut {NoStop}%
\bibitem [{\citenamefont {Fehske}\ and\ \citenamefont
  {Schneider}(2008)}]{Fehske08}%
  \BibitemOpen
  \bibfield  {author} {\bibinfo {author} {\bibfnamefont {H.}~\bibnamefont
  {Fehske}}\ and\ \bibinfo {author} {\bibfnamefont {R.}~\bibnamefont
  {Schneider}},\ }\href {http://dx.doi.org/10.1007/978-3-540-74686-7} {\emph
  {\bibinfo {title} {Computational many-particle physics}}}\ (\bibinfo
  {publisher} {Springer},\ \bibinfo {address} {Germany},\ \bibinfo {year}
  {2008})\BibitemShut {NoStop}%
\bibitem [{\citenamefont {Naldesi}\ \emph {et~al.}(2016)\citenamefont
  {Naldesi}, \citenamefont {Ercolessi},\ and\ \citenamefont
  {Roscilde}}]{Naldesi16}%
  \BibitemOpen
  \bibfield  {author} {\bibinfo {author} {\bibfnamefont {P.}~\bibnamefont
  {Naldesi}}, \bibinfo {author} {\bibfnamefont {E.}~\bibnamefont {Ercolessi}},
  \ and\ \bibinfo {author} {\bibfnamefont {T.}~\bibnamefont {Roscilde}},\
  }\href {\doibase 10.21468/SciPostPhys.1.1.010} {\bibfield  {journal}
  {\bibinfo  {journal} {SciPost Phys.}\ }\textbf {\bibinfo {volume} {1}},\
  \bibinfo {pages} {010} (\bibinfo {year} {2016})}\BibitemShut {NoStop}%
\bibitem [{\citenamefont {Khemani}\ \emph {et~al.}(2017)\citenamefont
  {Khemani}, \citenamefont {Sheng},\ and\ \citenamefont {Huse}}]{Khemani17}%
  \BibitemOpen
  \bibfield  {author} {\bibinfo {author} {\bibfnamefont {V.}~\bibnamefont
  {Khemani}}, \bibinfo {author} {\bibfnamefont {D.~N.}\ \bibnamefont {Sheng}},
  \ and\ \bibinfo {author} {\bibfnamefont {D.~A.}\ \bibnamefont {Huse}},\
  }\href {\doibase 10.1103/PhysRevLett.119.075702} {\bibfield  {journal}
  {\bibinfo  {journal} {Phys. Rev. Lett.}\ }\textbf {\bibinfo {volume} {119}},\
  \bibinfo {pages} {075702} (\bibinfo {year} {2017})}\BibitemShut {NoStop}%
\bibitem [{\citenamefont {{Guo}}\ \emph {et~al.}(2019)\citenamefont {{Guo}},
  \citenamefont {{Cheng}}, \citenamefont {{Sun}}, \citenamefont {{Song}},
  \citenamefont {{Li}}, \citenamefont {{Wang}}, \citenamefont {{Ren}},
  \citenamefont {{Dong}}, \citenamefont {{Zheng}}, \citenamefont {{Zhang}},
  \citenamefont {{Mondaini}}, \citenamefont {{Fan}},\ and\ \citenamefont
  {{Wang}}}]{Guo19}%
  \BibitemOpen
  \bibfield  {author} {\bibinfo {author} {\bibfnamefont {Q.}~\bibnamefont
  {{Guo}}}, \bibinfo {author} {\bibfnamefont {C.}~\bibnamefont {{Cheng}}},
  \bibinfo {author} {\bibfnamefont {Z.-H.}\ \bibnamefont {{Sun}}}, \bibinfo
  {author} {\bibfnamefont {Z.}~\bibnamefont {{Song}}}, \bibinfo {author}
  {\bibfnamefont {H.}~\bibnamefont {{Li}}}, \bibinfo {author} {\bibfnamefont
  {Z.}~\bibnamefont {{Wang}}}, \bibinfo {author} {\bibfnamefont
  {W.}~\bibnamefont {{Ren}}}, \bibinfo {author} {\bibfnamefont
  {H.}~\bibnamefont {{Dong}}}, \bibinfo {author} {\bibfnamefont
  {D.}~\bibnamefont {{Zheng}}}, \bibinfo {author} {\bibfnamefont {Y.-R.}\
  \bibnamefont {{Zhang}}}, \bibinfo {author} {\bibfnamefont {R.}~\bibnamefont
  {{Mondaini}}}, \bibinfo {author} {\bibfnamefont {H.}~\bibnamefont {{Fan}}}, \
  and\ \bibinfo {author} {\bibfnamefont {H.}~\bibnamefont {{Wang}}},\
  }\href@noop {} {\bibfield  {journal} {\bibinfo  {journal} {arXiv e-prints}\
  ,\ \bibinfo {eid} {arXiv:1912.02818}} (\bibinfo {year} {2019})},\ \Eprint
  {http://arxiv.org/abs/1912.02818} {arXiv:1912.02818 [quant-ph]} \BibitemShut
  {NoStop}%
\end{thebibliography}

\end{document}